\documentclass[prd]{revtex4}
\usepackage{graphicx}
\usepackage{bm}
\usepackage{diagbox}
\usepackage{amsmath}
\usepackage{array}
\usepackage{color}
\usepackage{eufrak}
\usepackage{mathrsfs}
\usepackage{ulem}
\usepackage{float}
\usepackage{caption}
\usepackage{color}
\usepackage{boxedminipage}
\usepackage[style]{fncychap}

\begin{document}
\title{Chiral Phase Transition in an Expanding Quark System}
\author{Ziyue Wang$^{1}$}
\email{zy-wa14@mails.tsinghua.edu.cn}
\author{Shuzhe Shi$^{2}$}
\author{Pengfei Zhuang$^{1}$}
\affiliation{$^{1}$ Physics Department, Tsinghua University, Beijing 100084, China}
\affiliation{$^{2}$ Department of Physics, McGill University, 3600 University Street, Montreal, QC, H3A 2T8, Canada}
\date{\today}
\begin{abstract}
We investigate the influence of chiral symmetry which varies along the space-time evolution of the system by considering the chiral phase transition in an non-equilibrium expanding quark-antiquark system. The chiral symmetry is described by the mean field order parameter, whose values is the solution of a self-consistent equation, and affects the space-time evolution of the system through the force term in the Vlasov equation. The Vlasov equation and the gap equation are solved concurrently and continuously for a longitudinal boost-invariant and transversely rotation-invariant system. This numerical framework enables us to carefully investigate how the phase transition and collision affect the evolution of the system. It is observed that the chiral phase transition gives rise to a kink in the flow velocity, which is caused by the force term in the Vlasov equation. The kink is enhanced by larger susceptibility and tends to be smoothed out by non-equilibrium effect. The spatial phase boundary appears as a ``wall'' for the quarks, as the quarks with low momentum are bounced back, while those with high momentum go through the wall but are slowed down.
\end{abstract}
\maketitle
\section{Introduction}
One of the major motivations in the study of finite-temperature Quantum ChromoDynamics~(QCD) is to shed light on the phase structure of the strong interaction matter. Lattice-QCD data has confirmed the chiral and deconfinement phase transition is a crossover for small baryon chemical potential~\cite{Aoki:2006we,Borsanyi:2010bp}. The sign problem at large baryon chemical potential~\cite{Fodor:2001au,deForcrand:2010ys} prevents lattice-QCD from giving precise predictions about the phase transition at finite density. Model calculations based on NJL model, quark meson model, and various beyond mean field frameworks have all predicted that the chiral phase transition at finite density is a first order phase transition~\cite{Scavenius:2000qd,Schaefer:2007pw,Fukushima:2008wg,Herbst:2010rf}. Thermodynamic theorem then predicts a critical end point~(CEP) between the crossover and the first order phase transition, which is a second order phase transition. However, due to various approximations adopted in the model calculations, there is not an agreement on the location of the CEP on the phase diagram. 

The exploration of the QCD phase diagram is also one of the most important goals for relativistic heavy-ion experiments. Through a systematic measurement over a range of beam energies, the beam energy scan~(BES) program makes it possible to search for the  CEP in the QCD phase diagram~\cite{Aggarwal:2010wy,Luo:2012kja,Adamczyk:2013dal}. Besides the ongoing BES program at RHIC, several other programs at other facilities such as FAIR and NICA have also contributed to the searching of QCD critical end point. The related experiments are mainly driven by measurements of net-proton or net-charge multiplicity fluctuations~\cite{Stephanov:1998dy,Stephanov:1999zu,Stephanov:2008qz} which are expected to show characteristic non-monotonic behavior near the phase transition and especially near the CEP~\cite{Stephanov:2011pb,Luo:2017faz}. The fast dynamics in the fireball renders it difficult to bridge the gap between the experiment and the theories. Since from the theoretical aspect, the QCD phase structure is investigated in an equilibrium, long-lived, extremely large and homogeneous system. On the contrary, the fireball in the heavy ion collision is a highly dynamical system, characterized by very short lifetime, extremely small size and fast dynamical expansion. The finite-size effect and off-equilibrium effect prevent a divergent correlation length, and thus weaken the critical phenomenon which takes place in the equilibrium system. On one hand, the fast expansion and cooling during the evolution of the fireball tends to drive the system out of local equilibrium. On the other hand, the relaxation time diverges around the critical point~\cite{Berdnikov:1999ph}, the critical slowing down renders it harder for the system to reach local equilibrium. A thorough understanding of phase transitions in the dynamical environment is thus of fundamental necessary to make profound predictions from the BES project. 

Various models have been applied to study the chiral phase transition and related critical phenomenon in an out-of-equilibrium system. In Ref.~\cite{Abada:1994mf, Abada:1996bw}, the authors investigate the chiral phase transition in a free-streaming quark-antiquark system by solving the Vlasov equation through the test particle method. The chiral fields are included considering their mean field values and their equations of motion. The Vlasov equation for the quark-antiquark system is also analytically solved in Ref.~\cite{Greiner:1996md}, assuming constant quark mass, and a shell-like structure at late evolution times in the center-of-mass (CM) frame is discovered. In Ref.~\cite{Yang:2003pz} the nonequilibrium and collision effects on the deconfinement phase transition is investigated by solving the Vlasov equation assuming Bjorken symmetry. The elastic two-body collisions for the quarks and antiquarks is included by simulating a Vlasov-type of equation with MonteCarlo test-particle approach~\cite{vanHees:2013qla,Meistrenko:2013yya,Wesp:2017tze}. Among the aforementioned works, although the force term is also considered in the test particle method~\cite{Abada:1994mf, Abada:1996bw,vanHees:2013qla,Meistrenko:2013yya,Wesp:2017tze}, its effect has not been carefully examined, which we find plays an important role in the evolution of the system around the phase transition.


In order to study the chiral phase transition in the non-equilibrium state, we investigate an expanding quark-anitquark gas system by solving the coupled Vlasov equation as well as the gap equation. This paper is arranged as follows. In section III, we introduce and analyze the coupled Vlasov equation and gap equation which we are going to solve, and give the related thermal quantities that can be obtained as momentum integrals of distribution function. In section III, we consider a longitudinal boost invariant and transverse rotational symmetric system, and derive the transport equation under such condition. In section IV, we present our numerical process to solve the coupled equations and then analyze the numerical result. In section V, we summarize this work and give a brief outlook.

\section{Vlasov equation and thermal quantities}

The partons in an off-equilibrium systems with background field and collisions can generally be described by the Vlasov equation
\begin{eqnarray}
\partial_t f^\pm \mp \boldsymbol{F} \cdot \boldsymbol{\nabla_p} f^\pm \pm \boldsymbol{v}\cdot \boldsymbol{\nabla_x} f^\pm = \mathcal{C}[f].
\end{eqnarray}
The above equation is applicable when the external fields and the interactions between the (quasi-)particles are sufficiently weak, so each particle can be considered to be moving along a classical trajectory, punctuated by rare collisions. As an example of an evolving global symmetry in an expanding parton system, we here consider the chiral symmetry in an off-equilibrium quark-antiquark system. At classical level, the velocity is $\boldsymbol{v}=\boldsymbol{p}/E_p$, the energy of the quasi-particle is $E_p=\sqrt{p^2+m(x)^2}$, where $m$ is the effective mass of the quark and antiquark, which is space-time dependent and is determined by the evolving chiral symmetry. The chiral mean field acts as a background field, and affects the motion of quarks through the gradient of the field energy $F=\boldsymbol{\nabla_x} E_p$, which is a continuous force on the quarks. The mass of the quasi-particles $m(x)$ is no longer a free parameter but is determined by the space-time dependent chiral symmetry. The constituent quark mass serves as the order parameter of chiral symmetry. Its temperature dependence in equilibrium  can be found in lattice-QCD simulation~\cite{Aoki:2006we,Borsanyi:2010bp} and other model calculations. We here consider the $SU(2)$ Nambu--Jona-Lasinio Lagrangian (NJL) model~\cite{Nambu:1961tp,Klevansky:1992qe,Hatsuda:1994pi},
\begin{equation}
\mathcal{L}=\bar{\psi}(i\gamma^\mu \partial_\mu-m_0)\psi+G\Big[(\bar{\psi}\psi)^2+(\bar{\psi}i\gamma_5{\bf \tau}\psi)^2\Big],
\end{equation}
where $\psi=(u,d)^T$ is the two-component quark field in the flavor space, $m_0$ is the degenerate current mass of the quarks, $\tau$ is the Pauli matrix in the isospin space. The transport equations can be derived from a first principle theory or an effective model in the framework of Wigner function~\cite{Vasak:1987um,BialynickiBirula:1991tx, Zhuang:1995pd,Zhuang:1995jb,Zhuang:1998bqx,Guo:2017dzf}. At the classical level, the quarks are treated as quasi-particles, and the chiral field is approximated by mean field, hence the Vlasov equation is coupled to the gap equation~\cite{Guo:2017dzf}. The disoriented chiral condensate (DCC)~\cite{Felder:2000hj,Cooper:1994ji,Gavin:1993bs} is negligible here because it appears as a quantum effect, and the $\sigma$ condensate appears as the mean field and couples to the Vlasov equation through the inhomogeneous quark mass. In order to investigate the collisions and non-equilibrium effect, we take the relaxation time approximation for the collision terms. The distribution function of the quark/antiquark number density $f^\pm(t,\boldsymbol{x},\boldsymbol{p})$ satisfies the coupled Vlasov equation and gap equation, 
\begin{eqnarray}
&& \partial_t f^\pm \mp \frac{\boldsymbol{\nabla_{r}}m^2}{2E_p} \cdot \boldsymbol{\nabla_p} f^\pm \pm \frac{\boldsymbol{p}}{E_p}\cdot \boldsymbol{\nabla_r} f^\pm = \mathcal{C}[f], \\
&& m \Big(1 + 2G \int \frac{\mathrm{d}^3\boldsymbol{p}}{(2\pi)^3} \frac{f^+(x,\boldsymbol{p})-f^-(x,\boldsymbol{p})}{E_p} \Big) = m_0,
\end{eqnarray}
where the $+$($-$) sign stands for quark (antiquark).
In this work, we take the relaxation time approximation for the collision term, $\mathcal{C}[f]=-(f^\pm-f^{\pm}_\mathrm{eq})/\tau_\theta$, with $f^{\pm}_\mathrm{eq}$ represents the corresponding local-equilibrium distribution function, while $\tau_\theta$ is the relaxation time. It is worth noticing that one can make the substitution $\widetilde{f}^-(t,\boldsymbol{x},\boldsymbol{p}) \equiv 1 - f^-(t,\boldsymbol{x},-\boldsymbol{p}) $ which follows the same equation of motion as $f^+$. The transport equations can be further simplified by adopting the recombination $f =  f^+ +  \widetilde{f}^-$ and $g =  f^+ - \widetilde{f}^-$. In such way, the evolution of $f$ and $g$ can be separated. Since $\widetilde{f}^-$ and $f^+$ satisfy the same transport equation, $f(t,\boldsymbol{x},\boldsymbol{p})$ and $g(t,\boldsymbol{x},\boldsymbol{p})$ also satisfy the same transport equation, while the gap equation depends only on $f$ but not $g$. One thus solve the coupled transport equation of distribution function $f(t,\boldsymbol{x},\boldsymbol{p})$ and $g(t,\boldsymbol{x},\boldsymbol{p})$ as well as the gap equation for a finite density system, and solve transport equation of distribution function $f(t,\boldsymbol{x},\boldsymbol{p})$ together with gap equation for a system with vanish baryon density,
\begin{eqnarray}
&& \partial_t f - \frac{\boldsymbol{\nabla_{r}}m^2}{2E_p} \cdot \boldsymbol{\nabla_p} f + \frac{\boldsymbol{p}}{E_p}\cdot \boldsymbol{\nabla_r} f = -\frac{f-f_\mathrm{eq}}{\tau_\theta}, \\
\label{transportf}
&& \partial_t g - \frac{\boldsymbol{\nabla_{r}}m^2}{2E_p} \cdot \boldsymbol{\nabla_p} g + \frac{\boldsymbol{p}}{E_p}\cdot \boldsymbol{\nabla_r} g = -\frac{g-g_\mathrm{eq}}{\tau_\theta}, \\
\label{gapeq}
&& m \Big(1 + 2G \int \frac{\mathrm{d}^3\boldsymbol{p}}{(2\pi)^3} \frac{f(x,\boldsymbol{p})}{E_p}- 2G \int \frac{\mathrm{d}^3\boldsymbol{p}}{(2\pi)^3} \frac{1}{E_p} \Big) = m_0.
\end{eqnarray}

\begin{figure}[H]\centering
\includegraphics[width=0.3\textwidth]{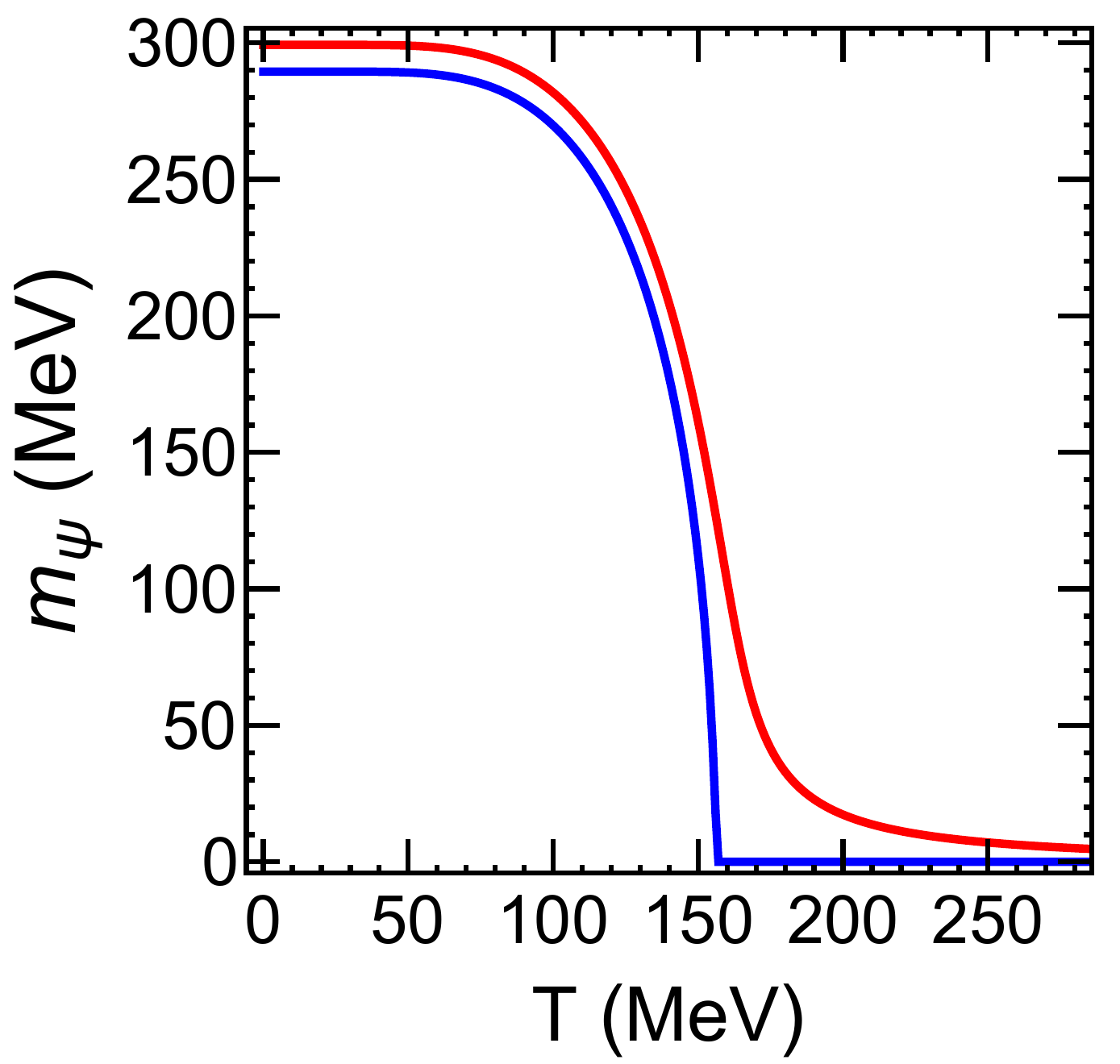}\qquad
\includegraphics[width=0.3\textwidth]{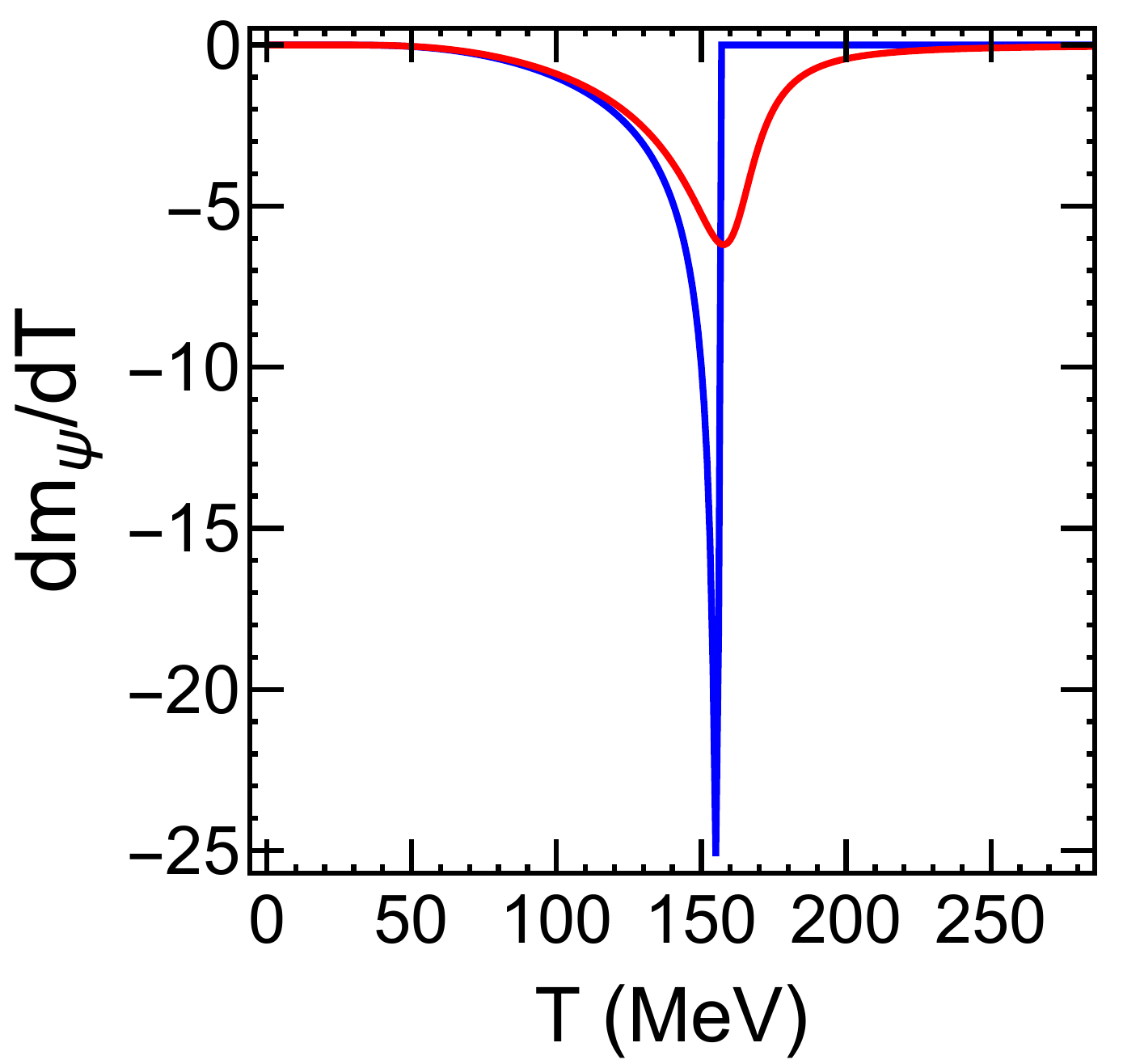}
\caption{Temperature dependence of quark mass in the real case and chiral limit for various baryon chemical potential in the equilibrium state.}
\label{phase transition}
\end{figure}

In the numerical procedures, we consider zero density system in this paper and solve both the transport equation of $f(t,\boldsymbol{x},\boldsymbol{p})$ and the gap equation, and eventually get the time and space dependence of the distribution function and the quark mass. The second integral in the gap equation is the vacuum part, which has ultra-violet divergence and needs to be regularized. Here we take the hard cut-off regularization only for the vacuum part,
\begin{eqnarray}
\int \frac{\mathrm{d}^3\boldsymbol{p}}{(2\pi)^3} \frac{1}{E_p} 
&=& 2\pi \int_0^{\Lambda}\int_0^{\Lambda} \frac{p_T\mathrm{d}p_T \mathrm{d}p_z}{(2\pi)^3} \frac{1}{E_p} \\
&=&  \frac{1}{4\pi^2} \bigg[\Lambda (\sqrt{m^2+2\Lambda^2} -\sqrt{m^2+\Lambda^2})
+ m^2 \ln\Big(\frac{m}{\Lambda+\sqrt{m^2+\Lambda^2}}\Big) + (m^2+\Lambda^2) \ln\Big(\frac{\Lambda+\sqrt{m^2+2\Lambda^2}}{\sqrt{m^2+\Lambda^2}}\Big) \bigg].\nonumber
\end{eqnarray}
The same cutoff $\Lambda=496$~MeV is adopted for the longitudinal momentum and the transverse momentum in the integral. The NJL coupling constant is set to be $G=1.688/\Lambda^2$, so as to guarantee the quark mass $m\sim 300$~MeV in the vacuum. The momentum integral of the finite temperature part is free from divergence, and is left unregularized. Under such choice of parameters, the temperature dependence of the quark mass in an equilibrium system can be directly calculated from the gap equation~(\ref{gapeq}) by taking Fermi-Dirac distribution. The quark mass for chiral limit $m_0=0$ and real case $m_0=3.7$~MeV are presented in Fig.~\ref{phase transition}, the critical temperature is about $156$~MeV at vanishing baryon chemical potential.
The chiral phase transition at vanishing baryon chemical potential is a crossover in the real case, and is a second order phase transition in the chiral limit. For a crossover, the phase transition can be defined at the maximum susceptibility $\mathrm{d}m/\mathrm{d}T$; for a second order phase transition, the susceptibility diverges at critical point. In an equilibrium system, the order parameter displays critical scaling $m\propto((T_c-T)/T_c)^\beta$ in the vicinity of a second order phase transition which can be described by critical exponent $\beta$. The divergence of the susceptibility is expected to affect the transport phenomenon of the system through the force term $\boldsymbol{\nabla_{r}}m^2 \cdot \boldsymbol{\nabla_p} f$, in which the force is provided by the ingredient of mass $\boldsymbol{\nabla_{r}}m^2$. In an equilibrium system with zero baryon density, the mass is determined by temperature alone, and the force can be expressed as $\boldsymbol{\nabla_{r}}m^2=2m(\mathrm{d}m/\mathrm{d}T)\boldsymbol{\nabla_{r}}T$. Since the susceptibility exhibits a peak around the phase transition, the temperature gradient $\boldsymbol{\nabla_{r}}T$ in a realistic system is nonzero, the force term is expected be very large at the phase transition point in the time-space. However, when the system is close to the second order phase transition, the relaxation time may diverge~\cite{Berdnikov:1999ph}, the critical slowing down takes place, makes it harder for the system to reach local equilibrium. When the system has not yet reached local equilibrium, the temperature is not well-defined, and neither for the expression $2m(\mathrm{d}m/\mathrm{d}T)\boldsymbol{\nabla_{r}}T$ of the force term. If the phase transition takes place in an out-of equilibrium system, the aforementioned effects of the force term may have been overestimated. In the following, we study the chiral phase transition in both local equilibrium and out-of-equilibrium systems, controlled by the relaxation time. Then we analyze the influence of the phase transition and the force term on the evolution of the system. 

The thermodynamic quantities can be constructed from the distribution function. For single component medium, positive particle $f^+$ for example, the current and energy-momentum stress tensor are defined by:
\begin{eqnarray}
J_+^\mu(t,\boldsymbol{x}) &=& \int \frac{\mathrm{d}^3\boldsymbol{p}}{(2\pi)^3E_p} p^\mu f^+(t,\boldsymbol{x},\boldsymbol{p}) , \nonumber \\
T_+^{\mu\nu} (t,\boldsymbol{x})&=& \int  \frac{\mathrm{d}^3\boldsymbol{p}}{(2\pi)^3E_p}p^\mu p^\nu f^+(t,\boldsymbol{x},\boldsymbol{p}),
\end{eqnarray}
which should satisfy that the conservation laws $\partial_\mu J^\mu=0$ and $\partial_\nu T^{\mu\nu}=0$. Since the total energy density and entropy is the sum of that of positive particles and negative particles, while the net-quark number density is the difference of positive particles and negative particles, the above equations can be rewritten using the redefined distribution function $f =  f^+ +  \widetilde{f}^-$ and $g =  f^+ - \widetilde{f}^-$,
\begin{eqnarray}
\begin{split}
J^\mu(t,\boldsymbol{x}) =& \int \frac{\mathrm{d}^3\boldsymbol{p}}{(2\pi)^3E_p} p^\mu g(t,\boldsymbol{x},\boldsymbol{p})  ,  \\
T^{\mu\nu} (t,\boldsymbol{x})=& \int  \frac{\mathrm{d}^3\boldsymbol{p}}{(2\pi)^3E_p}p^\mu p^\nu f(t,\boldsymbol{x},\boldsymbol{p}) , \\
S^\mu(t,\boldsymbol{x})=&-\int\frac{d^3\boldsymbol{p}}{(2\pi)^3E_p}p^\mu
	\Big(f^+(t,\boldsymbol{x},\boldsymbol{p})\ln f^+(t,\boldsymbol{x},\boldsymbol{p})+(1-f^+(t,\boldsymbol{x},\boldsymbol{p}))\ln(1-f^+(t,\boldsymbol{x},\boldsymbol{p})) \\
	&\qquad\qquad\qquad\;\;+f^-(t,\boldsymbol{x},\boldsymbol{p})\ln f^-(t,\boldsymbol{x},\boldsymbol{p})+(1-f^-(t,\boldsymbol{x},\boldsymbol{p}))\ln(1-f^-(t,\boldsymbol{x},\boldsymbol{p}))
	\Big).
\end{split}
\end{eqnarray}
Using the Landau frame definition, the fluid velocity can be determined as the {\it time-like} eigenvector ($u^\mu u_\mu >0$) of the stress tensor $T^{\mu}_{\;\;\nu} u^\nu = \epsilon\; u^\mu$, with energy density $\epsilon$ being the corresponding eigenvalue. One could further obtain the particle number density and entropy density as $n =u_\mu J^\mu$ and $s=S^\mu u_\mu$. The time-space evolution of $\epsilon(t,\textbf{x})$, $n(t,\textbf{x})$ and $u^z(t,\textbf{x})$ would give us quantitative idea about how the system evolves.

By taking the relaxation time approximation for collision kernel, we also need the corresponding local-equilibrium distribution function for any given time and space point,
\begin{equation}
f^\pm_\mathrm{eq}(x,\boldsymbol{p})=\frac{1}{e^{(\pm u_\mu p^\mu + \mu_\mathrm{eq})/T_\mathrm{eq}}+1},
\label{feq}
\end{equation}
In the above distribution, the temperature and chemical potential are determined by matching the energy and number density, i.e. $\epsilon = \epsilon_\mathrm{eq}$ and $n = n_\mathrm{eq}$, where
\begin{eqnarray}
 \epsilon_\mathrm{eq} &\equiv&  
	\int \frac{ (u \cdot p)^2\mathrm{d}^3\boldsymbol{p}}{(2\pi)^3E_p} 
		f_\mathrm{eq}(t,\boldsymbol{x},\boldsymbol{p}) \,, \\
n_\mathrm{eq} &\equiv&
	\int \frac{ (u \cdot p)\mathrm{d}^3\boldsymbol{p}}{(2\pi)^3E_p} 
		g_\mathrm{eq}(t,\boldsymbol{x},\boldsymbol{p}) \,,
\end{eqnarray}
with $E_p=\sqrt{m(T_\mathrm{eq},\mu_\mathrm{eq})^2+p^2}$.
Similarly, we define the equilibrium limit of the entropy density as
\begin{equation}
s_\mathrm{eq}=-\int\frac{(u \cdot p) d^3\boldsymbol{p}}{(2\pi)^3E_p} \Big(
	f_\mathrm{eq}^+ \ln f_\mathrm{eq}^+ +(1-f_\mathrm{eq}^+)\ln(1-f_\mathrm{eq}^+)
	+f_\mathrm{eq}^- \ln f_\mathrm{eq}^- +(1-f_\mathrm{eq}^-)\ln(1-f_\mathrm{eq}^-)
	\Big) .
\end{equation}
The difference between $s_\mathrm{eq}$ and the actual entropy density $s$ quantifies how close the system is to the equilibrium state.
In a zero chemical potential limit considered in this paper, one would automatically find $g=0$, hence $J=0$, $n=n_\mathrm{eq}=0$, $\mu_\mathrm{eq}$ = 0.

\section{Symmetry and Simplification}
For numerical simplicity, we will focus on the {\it longitudinal boost-invariant} and {\it transversal rotational-symmetric} systems, which is a good approximation for ultra-central relativistic heavy-ion collisions. Under such symmetries, two constraints are applied to the system, and the distribution function $f$ five degrees of freedom. We introduce a new set of coordinates $(\tau,\eta,\rho,\phi, p_\perp,\xi,\theta)$, the original coordinates and the new ones can be transformed through the following relation,
\begin{eqnarray}
t &=& \tau \cosh\eta, \qquad p_t~=~\sqrt{m(\tau,\rho)^2+p_\perp^2}\cosh(\xi+\eta),\nonumber\\
z &=& \tau \sinh\eta, \qquad p_z~=~\sqrt{m(\tau,\rho)^2+p_\perp^2}\sinh(\xi+\eta),\nonumber\\
x &=& \rho \cos\phi, \qquad~ p_x~=~p_\perp\cos(\phi+\theta),\nonumber\\
y &=& \rho \sin\phi, \qquad~ p_y~=~ p_\perp \sin (\phi+\theta),
\end{eqnarray}
where $\rho\in[0,+\infty)$, $p_\perp\in[0,+\infty)$ and $\xi \in(-\infty,+\infty)$. Under the new set of coordinate, the longitudinal boost-invariance and transversal rotational-symmetry of the distribution function can be translated into its independence of $\phi$ and $\eta$, namely $\partial_\phi f=\partial_\eta f=0$. The phase space of the distribution function becomes $(\tau,\rho, p_\perp,\xi,\theta)$. The transport equation~(\ref{transportf}) is then reduced to 
\begin{eqnarray}
\partial_\tau f
+\frac{p_\perp\cos\theta}{E_p}\partial_\rho f
-\frac{m(\partial_\rho m)\cos\theta}{E_p}\partial_{p_\perp} f
-\tanh\xi\left(\frac{1}{\tau}+\frac{m(\partial_\tau m)}{p_\perp^2+m^2}\right)\partial_\xi f
-\frac{\sin\theta}{E_p}\left(\frac{p_\perp}{\rho}-\frac{m(\partial_\rho m)}{p_\perp}\right)\partial_\theta f
=-\frac{f-f_\mathrm{eq}}{\tau_\theta},
\label{transport_gubser}
\end{eqnarray}
where $f_\mathrm{eq}$ is the equilibrium distribution.
To clearly analyze the $\theta$-dependence of the distribution function, and to simplify the calculation, we take the Fourier expansion of $f(\tau,\rho, p_\perp,\xi,\theta)$ with respect to $\theta$, and also the equilibrium distribution function $f_\mathrm{eq}(\tau,\rho, p_\perp,\xi,\theta)$,
\begin{eqnarray}
\begin{split}
f(\tau,\rho, p_\perp,\xi,\theta) =&\; a_0(\tau,\rho, p_\perp,\xi)+2\sum_{n=1}^{\infty}\left[a_n(\tau,\rho, p_\perp,\xi)\cos(n\theta)+b_n(\tau,\rho, p_\perp,\xi)\sin(n\theta)\right],\\
f_\mathrm{eq}(\tau,\rho, p_\perp,\xi,\theta) =&\; A_0(\tau,\rho, p_\perp,\xi)+2\sum_{n=1}^{\infty}\left[A_n(\tau,\rho, p_\perp,\xi)\cos(n\theta)+B_n(\tau,\rho, p_\perp,\xi)\sin(n\theta)\right],
\end{split}\label{fourier}
\end{eqnarray}
where the Fourier coefficients are obtained by definition $a_n(\tau,\rho, p_\perp,\xi)\equiv (2\pi)^{-1}\int_{-\pi}^\pi f(\tau,\rho, p_\perp,\xi,\theta)\cos(n \theta) \mathrm{d}\theta$ and $b_n(\tau,\rho, p_\perp,\xi)\equiv(2\pi)^{-1}\int_{-\pi}^\pi f(\tau,\rho, p_\perp,\xi,\theta)\sin(n \theta) \mathrm{d}\theta$, and similarly for $A_n$ and $B_n$.
In a realistic system, one can further expect its symmetry under reflection along either $\boldsymbol{\hat{x}}$-, $\boldsymbol{\hat{y}}$-, or $\boldsymbol{\hat{z}}$-direction. Under such condition, one can show the $\theta$-odd components of $f$ and $f_\mathrm{eq}$ vanish, $b_n \equiv B_n \equiv 0$, as well as $a_n(-\xi) = a_n(\xi)$, $A_n(-\xi) = A_n(\xi)$. Substituting the Fourier expansion~(\ref{fourier}) back into the Vlasov equation~(\ref{transport_gubser}), we then reduce the Vlasov equation to the transport equations of the corresponding Fourier components $a_n$, 
\begin{eqnarray}
\label{gubser_couple}
\begin{split}
&(\partial_\tau a_0)-\tanh\xi\left(\frac{1}{\tau}+\frac{m(\partial_\tau m)}{p_\perp^2+m^2}\right)(\partial_\xi a_0)+\frac{p_\perp}{E_p}(\partial_\rho a_1)-\frac{m(\partial_\rho m)}{E_p}(\partial_{p_\perp} a_1)+\frac{1}{E_p}\left(\frac{p_\perp}{\rho}-\frac{m(\partial_\rho m)}{p_\perp}\right)a_1
=-\frac{a_0-A_0}{\tau_\theta},\\
&(\partial_\tau a_n)-\tanh\xi\left(\frac{1}{\tau}+\frac{m(\partial_\tau m)}{p_\perp^2+m^2}\right)(\partial_\xi a_n)+\frac{p_\perp}{2E_p}\partial_\rho\Big( a_{n-1}+ a_{n+1}\Big)
-\frac{m(\partial_\rho m)}{2E_p}\partial_{p_\perp}\Big( a_{n-1}+a_{n+1}\Big)\\
&\qquad\qquad\qquad\qquad\qquad\qquad\qquad\qquad~~~~
-\frac{1}{2E_p}\left(\frac{p_\perp}{\rho}-\frac{m(\partial_\rho m)}{p_\perp}\right)\Big((n-1)a_{n-1}-(n+1)a_{n+1}\Big)
=-\frac{a_n-A_n}{\tau_\theta},
\end{split}
\label{transport2}
\end{eqnarray}
where the energy of quasi-particle is $E_p(\tau,\rho)=\sqrt{m^2(\tau,\rho)+p_\perp^2}\cosh(\xi)$, with mass $m(\tau,\rho)$ obtained from the gap equation, which in the new coordinate becomes,
\begin{eqnarray}
 m \Big(1 + 2N_dG \int \frac{p_\perp\mathrm{d}p_\perp\mathrm{d}\xi}{2(2\pi)^2} 
\big(a_0(\tau,\rho, p_\perp,\xi)-1\big)\Big) = m_0.
\end{eqnarray}
The momentum integral in the gap equation only relates to the zeroth component $a_0$, while the first and second order Fourier components contribute to the currents and energy-momentum tensor. 

The energy-momentum tensor is 
\begin{eqnarray}
T^{\mu}_{\;\;\nu} = \left( \begin{array}{cccc}
T^{\tau\tau} & 0 & - T^{\tau\rho} & 0 \\
0 & -\tau^2T^{\eta\eta} & 0 & 0 \\
T^{\rho\tau} & 0 & -T^{\rho\rho} & 0 \\
0 & 0 & 0 & -\rho^2T^{\phi\phi} \\
\end{array}\right),
\end{eqnarray}
with the non-vanishing components are
\begin{eqnarray}
\begin{split}
T^{\tau\tau}(\tau,\rho) =& ~~~\int_pa_0(\tau,\rho, p_\perp,\xi) \left(m(\tau,\rho)^2+p_\perp^2\right)\cosh^2(\xi),  \\
T^{\eta\eta}(\tau,\rho) =&\frac{1}{\tau^2}\int_p a_0(\tau,\rho, p_\perp,\xi) \left(m(\tau,\rho)^2+p_\perp^2\right)\sinh^2(\xi),  \\
T^{\tau\rho}(\tau,\rho) =&~~~\int_pa_1(\tau,\rho, p_\perp,\xi) p_\perp\sqrt{m(\tau,\rho)^2+p_\perp^2}\cosh(\xi),  \\
T^{\rho\rho}(\tau,\rho) =&~~~\int_p\left(a_0(\tau,\rho, p_\perp,\xi) +a_2(\tau,\rho, p_\perp,\xi)\right)p^2_\perp/2,\\
T^{\phi\phi}(\tau,\rho) =&\frac{1}{\rho^2}\int_p \left(a_0(\tau,\rho, p_\perp,\xi) -a_2(\tau,\rho, p_\perp,\xi)\right)p^2_\perp/2,
\end{split}
\label{tensor}
\end{eqnarray}
where $\int_p$ is the abbreviation for $\int \frac{p_\perp\mathrm{d}p_\perp\mathrm{d}\xi}{4(2\pi)^2}$. For this energy-momentum tensor, one can explicitly write down the flow velocity and energy density, being the time-lime eigenvector and the corresponding eigenvalue:
\begin{eqnarray}
\begin{split}
\epsilon&=\Big(T^{\tau\tau}-T^{\rho\rho}+\sqrt{(T^{\tau\tau}+T^{\rho\rho})^2-4(T^{\tau\rho})^2}\Big)/2,\\
u^\mu&\equiv \{u^\tau, 0, u^\rho, 0\}= 
\bigg\{\Big( \frac{T^{\tau\tau}+T^{\rho\rho}}{2\sqrt{(T^{\tau\tau}+T^{\rho\rho})^2-4(T^{\tau\rho})^2}} + \frac{1}{2}\Big)^{1/2},0,
\Big( \frac{T^{\tau\tau}+T^{\rho\rho}}{2\sqrt{(T^{\tau\tau}+T^{\rho\rho})^2-4(T^{\tau\rho})^2}} - \frac{1}{2}\Big)^{1/2},
0\bigg\},
\label{eigen}
\end{split}
\end{eqnarray}
Noting that the velocity has only two non-zero components, $u^\tau$ and $u^\rho$, the equilibrium distribution function~(\ref{feq}) can be expressed as
\begin{eqnarray}
f_\mathrm{eq}(\tau,\rho, p_\perp,\xi,\theta)=\frac{1}{\exp\left(T_\mathrm{eq}^{-1} u_\tau \sqrt{m^2+p_\perp^2}\cosh\xi - T_\mathrm{eq}^{-1} u_\rho p_\perp\cos\theta\right)+1}.
\end{eqnarray}

\section{Numerical Procedure and Result}
In the numerical procedures, we solve the finite difference versions of transport equations. The distribution function is discretized on a fixed grid in the calculation frame. The phase space is discretized as follows, take 200 points for $\rho$ within the range $\rho/\rho_0\in[-3,3]$, 100 points for $p_T$ within $p_T/T_0\in[0,8]$, 100 points for $\xi$ within the range $\xi\in[0,6]$. The Fourier expansion of distribution function $f$ with respect to $\theta$ is taken with maximum $n=7$ to guarantee the convergence. To eschew the numerical instability around $\tau=0$, we take the initial time as $\tau_0=0.5$~fm, the time step in the evolution is taken to be $\mathrm{d}\tau=0.0005$~fm to guarantee the stability. The calculation at the discrete time step n+1 involves only quantities at the previous time step. At each time step, we solve both the transport equation and the gap equation, and eventually get the time and space dependence of the distribution function and the quark mass. This numerical framework is verified to be reliable by comparing the result with the analytical solution in a spherical symmetric system~\cite{Greiner:1996md}, as well as checking the conservation of the particle number and energy-momentum.

The initial state of the fireball is a highly off-equilibrium system, the evolution towards local equilibrium quark gluon plasma is also an interesting problem. However, this is not our concern in this paper, since the temperature is not well-defined in such initial stage, the discussion of phase transition is also questionable. We here discuss the evolution of the system from local equilibrium towards off-equilibrium state after the formation of quark gluon plasma. A local equilibrium initial state is adopted, and a local temperature could be assigned. The large gradient in the initial condition drives towards off-equilibrium state, while the collisions drives the system towards local equilibrium. In order to describe the hot chiral restored medium in the inner part and the cold chiral symmetry broken medium in the outer part, we choose a Gaussian temperature profile $T(\rho)=T_0\exp\left(-\rho^2/\rho_0^2\right)$ for the initial state,  with $T_0=300$~MeV and $\rho_0=2$~fm. For the local equilibrium initial state, the distribution function is the Fermi-Dirac distribution, $a_0(\tau_0, \rho,  p_\perp,\xi)=2(e^{E_p/T(\rho)}+1)^{-1}$, where energy is $E_p=\sqrt{m^2+p_\perp^2}\cosh\xi$, and $T(\rho)$ is the above initial temperature profile. One can easily check that all other Fourier components vanish, $a_i (\tau_0)=0$ for $i\geq 1$. In the real case, the current mass is chosen to be $m_0=3.7$~MeV, and in the chiral limit, we have the current mass $m_0=0$.

The chiral phase transition is characterized by the chiral order parameter $\sigma$, or the quark constituent mass. The constituent mass is generated by the gap equation at each space-time point, and enters the transport equation through three ways: the energy $E_p=\sqrt{m^2+p_\perp^2}\cosh\xi$, the evolution rate of constituent mass $\partial_\tau m$, and through the spatial gradient of mass $\partial_\rho m$. In order to illustrate the influence of the phase transition and the force term in the transport equation, we consider the following three different conditions. First, when solving the transport equation and the gap equation concurrently, the effect of the phase transition and the force terms are both taken into consideration. Second, for a comparison, we solve the transport equation alone and keep the quark mass as a constant, for instance $m=150$~MeV. In this case, there is no phase transition nor force term. Third, in order to further illustrate the influence of the force term, we solve the transport equation and the gap equation at each step, but ignore the force term in the transport equation, namely assuming $\partial_\tau m=0$ and $\partial_\rho m=0$. We also study the influence of out-of-equilibrium effect by comparing the results of different relaxation time, for small relaxation time the system stays close to local equilibrium; while for large relaxation time, the system is away from equilibrium.

\subsection{Thermodynamical quantities}
\begin{figure}[H]\centering
\includegraphics[width=0.35\textwidth]{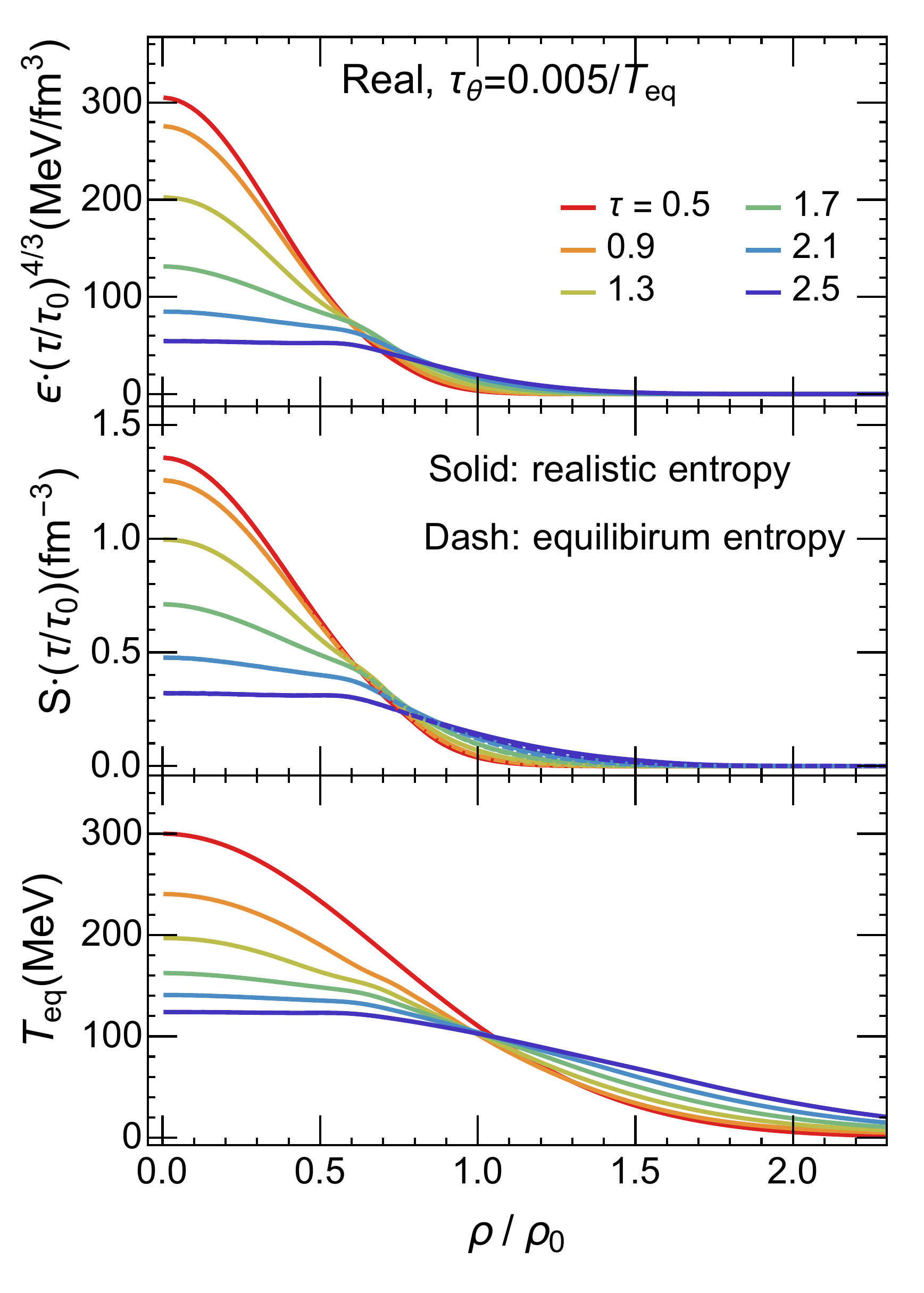}\qquad
\includegraphics[width=0.35\textwidth]{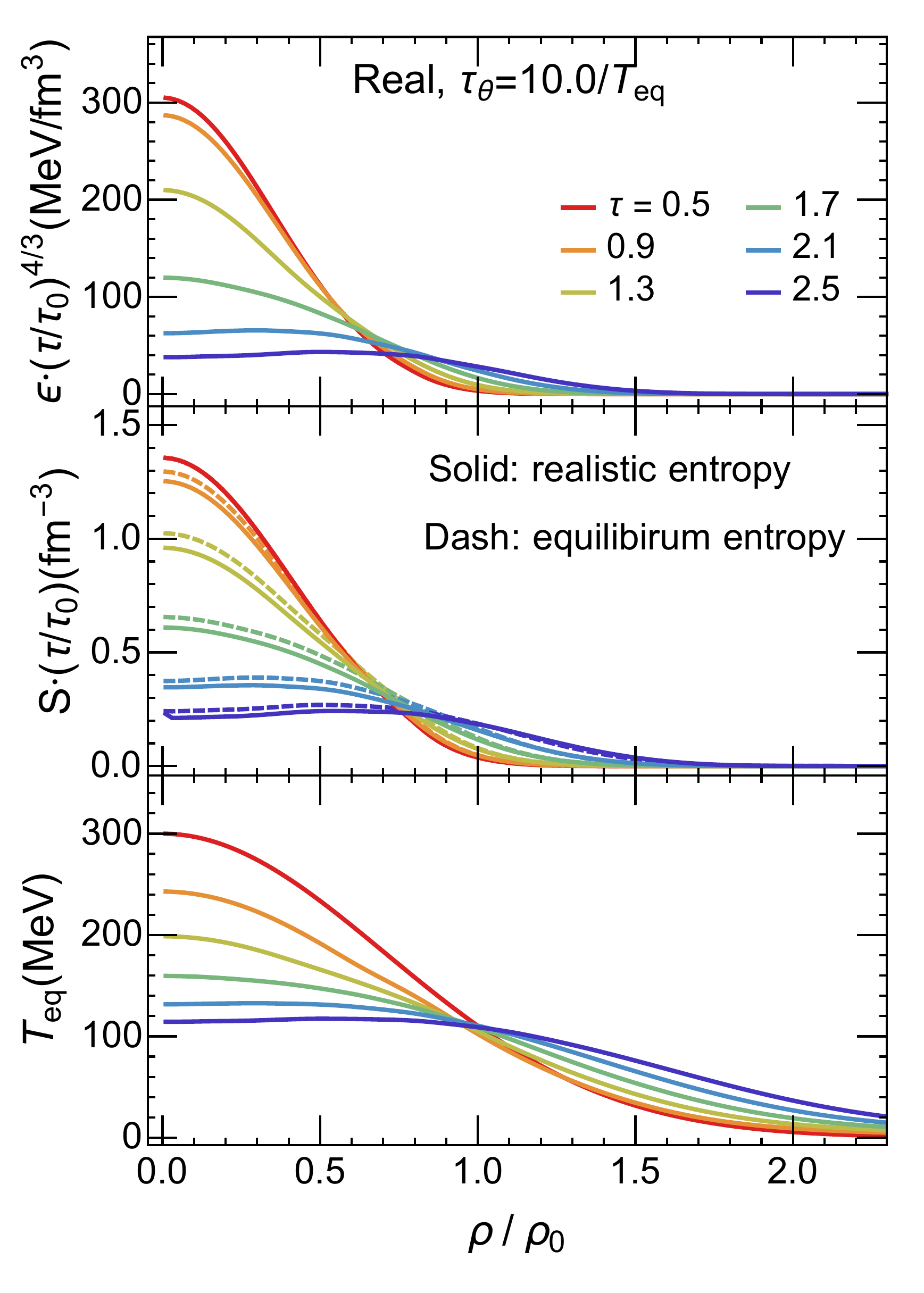}
\caption{The energy density (top), entropy density (middle) and equilibrium temperature (bottom) in the expanding system. The left and right panel corresponds to small and large relaxation time. The lines are rainbow colored, representing different evolution time. In the figure of entropy density, the solid and dashed lines correspond to realistic and equilibrium entropy.}
\label{gub_thermal}
\end{figure}
First, we self-consistently solve the coupled transport equation as well as the gap equation with finite current mass, and adopt different relaxation time. At each time step, the distribution function is obtained by evolving the transport equation, the energy density and entropy density are calculated by definition Eq.~(\ref{current}) and Eq.~(\ref{tensor}). The temperature is obtained by matching the realistic energy density to that of the equilibrium state. The energy density, entropy and temperature are presented in Fig.~\ref{gub_thermal}. The left panel corresponds to the evolution with small relaxation time, and the right panel corresponds to those of large relaxation time. The lines are rainbow colored which represents different evolution time, from the red line to the purple line represent the initial distribution to the distribution at later time. 

The initial condition of the system is chosen to be an equilibrium distribution, with a gaussian distribution temperature profile, the inner part (small $\rho$) has higher temperature and the outer part (large $\rho$) has lower temperature. With the expansion of the system, the temperature of the core area gradually decrease, and the temperature of the outer area increases. The initial large gradient of the energy density drives the system away from the local equilibrium, while the collisions bring the system back to equilibrium. The entropy tells whether the system has reached local equilibrium, the solid line represent the realistic entropy density and the dashed lines represent the entropy of the equilibrium state. Since the equilibrium state takes the maximum entropy. When the collisions are not strong enough, the system takes longer time in the out-of-equilibrium state, where the realistic entropy is smaller than that of the equilibrium state. If the relaxation time is small,
the collisions are strong enough to keep the system at local equilibrium, the entropy density of the state is the same as that of the equilibrium state.

\subsection{Constituent mass and Phase boundary}
\begin{figure}[H]\centering
\includegraphics[width=0.35\textwidth]{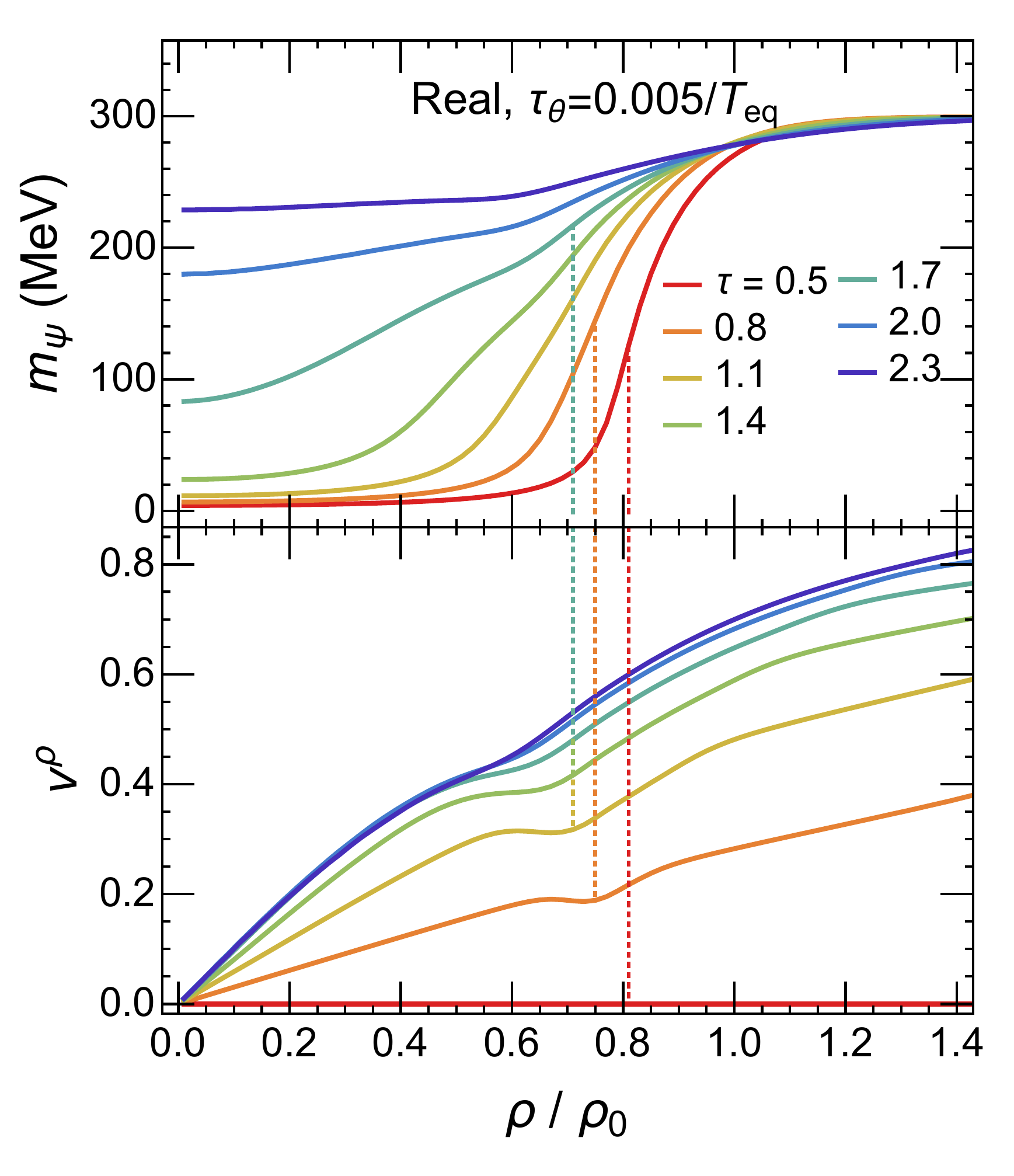}
\includegraphics[width=0.35\textwidth]{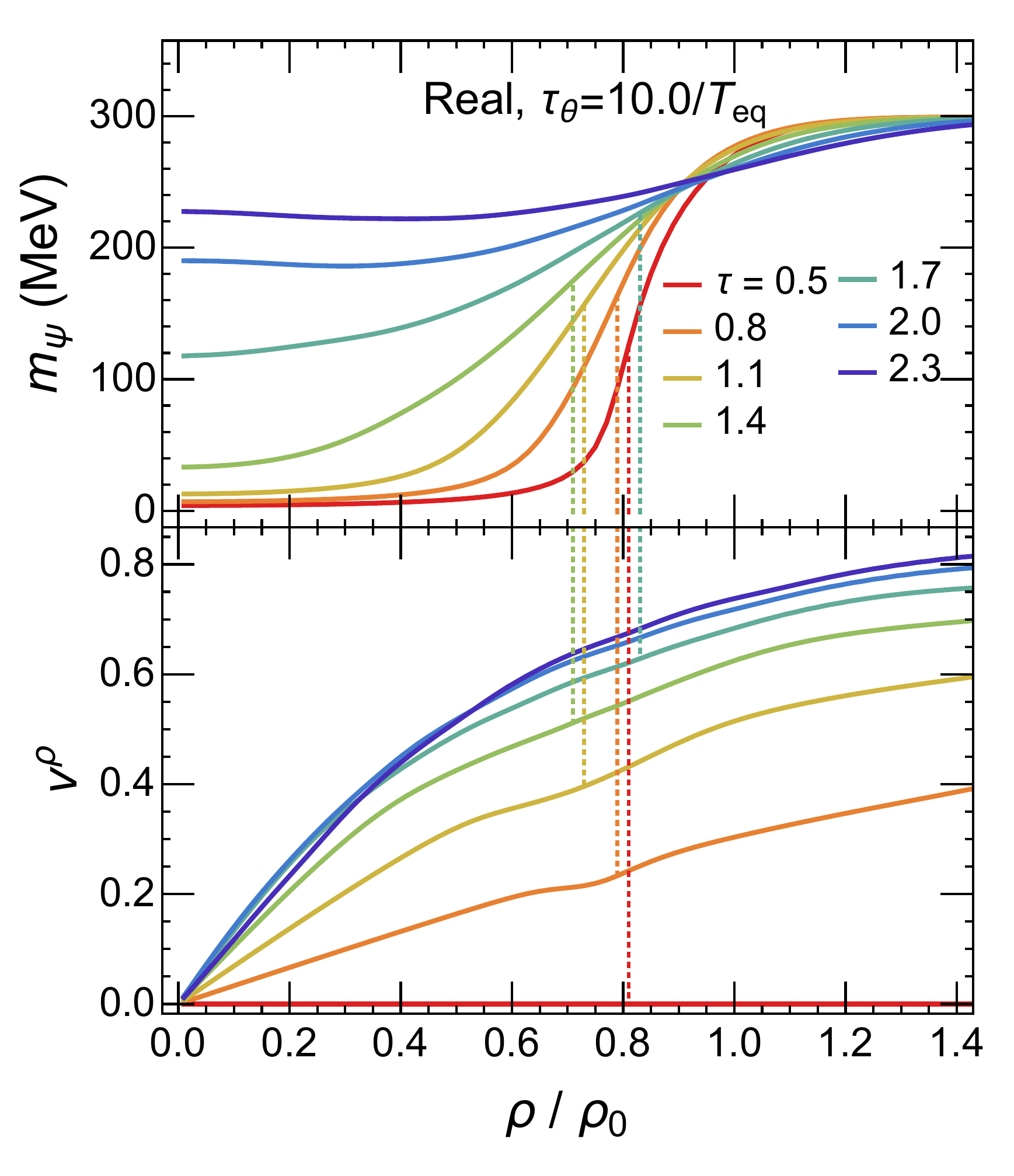}\\
\includegraphics[width=0.35\textwidth]{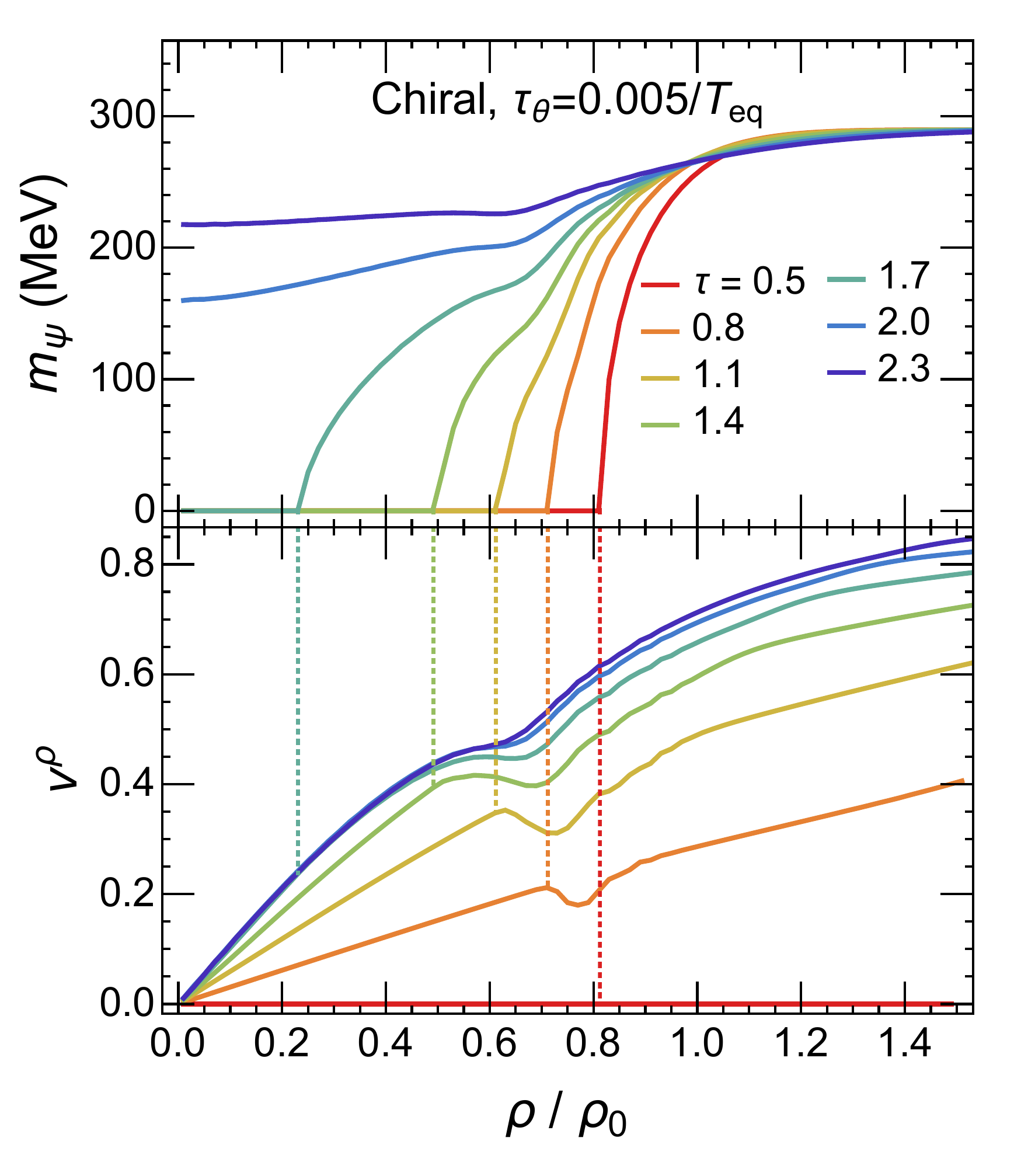}
\includegraphics[width=0.35\textwidth]{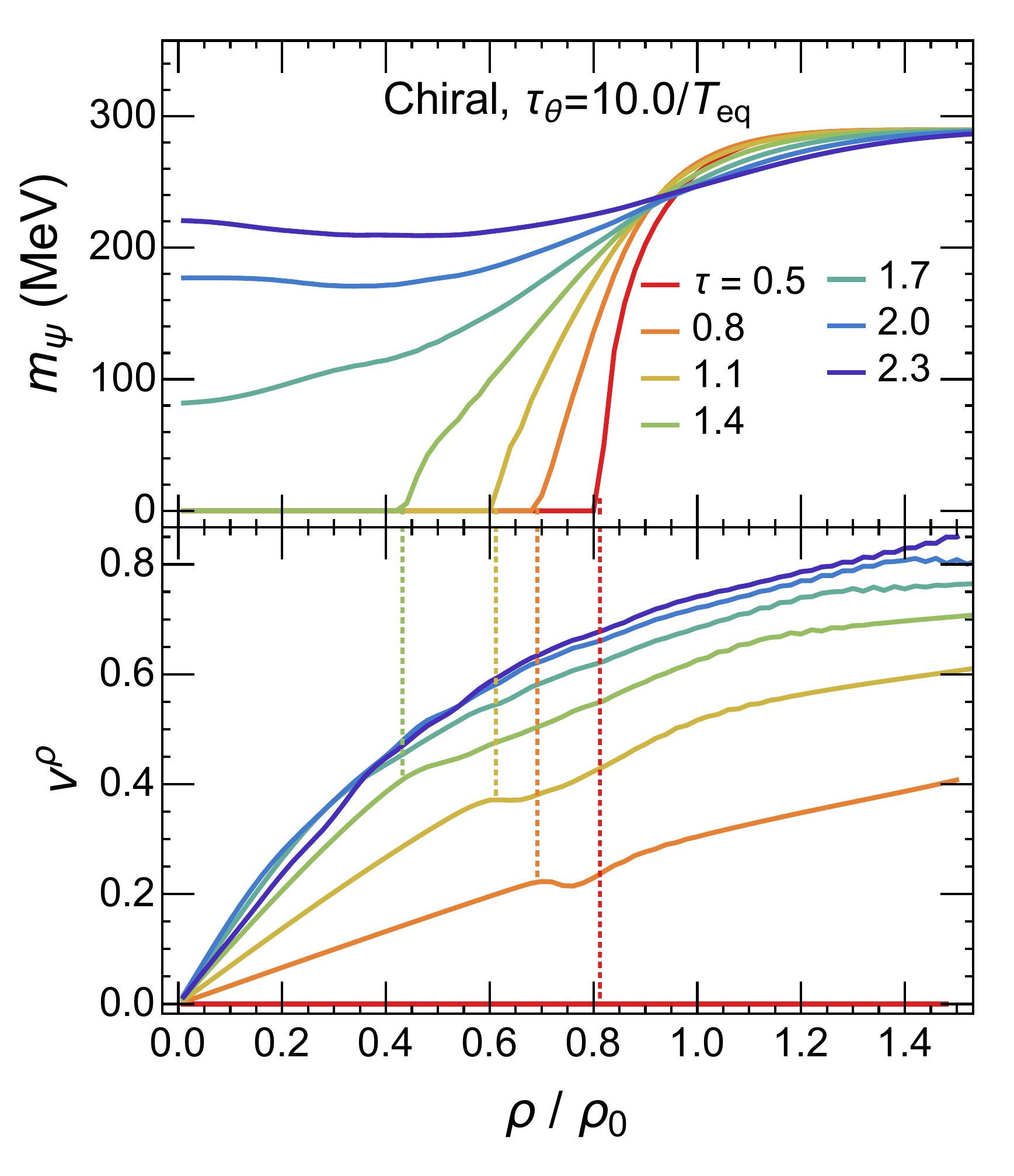}
\caption{The evolution of quark mass $m_\psi(\tau,\rho)$ and transverse velocity $v_\rho(\tau,\rho)$ in the real case and chiral limit, with a comparison between large and small relaxation time. The lines are rainbow colored representing different evolution times. The phase transition point is marked out by the dotted line.}
\label{gub_mass}
\end{figure}
In the equilibrium state, the quark constituent mass $m_\psi$ serves as the order parameter of the chiral symmetry, which tells to what extent the chiral symmetry is broken. Here in an expanding quark-antiquark system, the space-time dependent quark mass $m_\psi(\tau,\rho)$ signals the chiral symmetry in the space-time. In the initial state, the inner part of the system has higher temperature and is in the chiral symmetry restored phase, the constituent mass is small; the outer part of the system has lower temperature and is in the chiral symmetry broken phase with large constituent mass. By solving together the coupled transport equation and gap equation, the evolution of constituent quark mass in the time-space can be obtained self-consistently. The evolution of both the real case and chiral limit are investigated for large and small relaxation time, the results are presented in Fig.~\ref{gub_mass}. 

The upper two figures present the quark mass $m_\psi(\tau,\rho)$ and transverse velocity $v_\rho(\tau,\rho)\equiv(u^\rho/u^\tau)$ in the real case, the lower two figures correspond to those of chiral limit. With the expansion of the system, the quark mass in the inner area grows with time, indicating the gradually restoring of chiral symmetry; the quark mass of outer area decreases, indicating the breaking of chiral symmetry. In the equilibrium chiral phase transition, the phase transition point in the chiral limit is well-defined, while that of crossover does not has a strict definition, one of the usually used definition is the maximum of susceptibility $\mathrm{d}m_\psi/\mathrm{d}T$. In the expanding system, the phase transition point in the chiral limit can still be defined by the time-space point where that quark mass reached zero. For the real case, we here take phase boundary as the space-time point where the equilibrium temperature is around the critical temperature and $\mathrm{d}m/\mathrm{d}x$ takes the maximum. The phase transition points are marked out by the dotted lines in the Fig.~\ref{gub_mass}.
\begin{figure}[H]\centering
\includegraphics[width=0.35\textwidth]{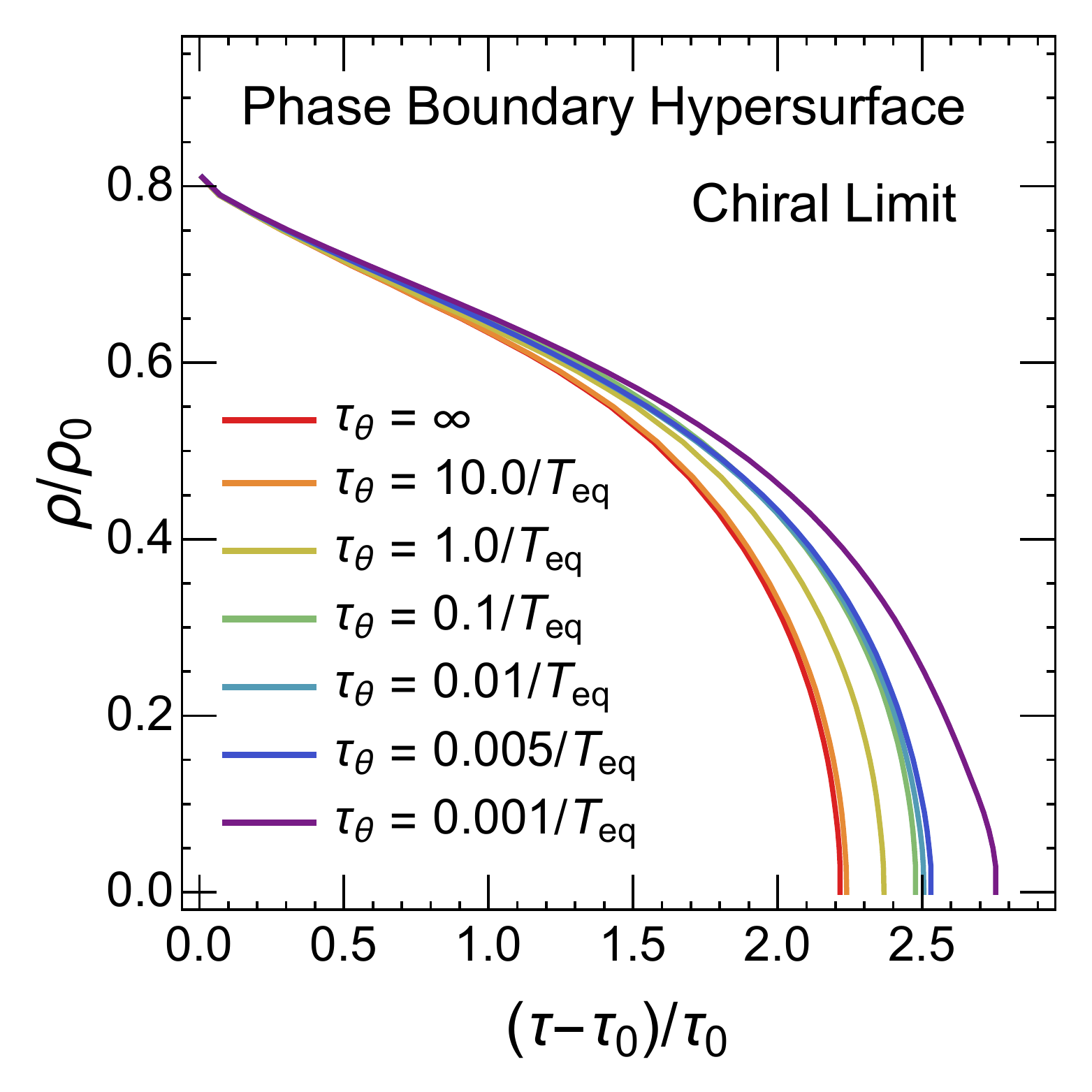}
\caption{The phase transition hypersurface obtained by self-consistent solution of a chiral limit system. Different lines represent different relaxation time.}
\label{gub_boundary}
\end{figure}
The phase diagram of equilibrium strong interaction matter in the $T-\mu$ plane is a map for the chiral symmetry. For non-equilibrium quark matter, the chiral symmetry breaking and restoration can be described by a phase diagram in space-time, with the phase boundary hypersurface indicates where the symmetry changes in the time-space. In the chiral limit, the phase transition hypersurface can still be defined by $x^\mu(m_\psi=0)$ no matter whether the system is in the local equilibrium or not. Fig.~\ref{gub_boundary} presents the hypersurface obtained by self-consistent solution of a chiral limit system, with different colored lines represent different relaxation time. The collisions cast the kinetic energy into internal energy, thus decelerates the expansion of the system. In the initial state $(\tau-\tau_0)/\tau_0=0$, the core area $\rho/\rho_0<0.8$ is in the chiral symmetry restored phase. With the expansion of the system, the core area gradually cools, the area of chiral symmetry restored phase shrinks. The free streaming system expands the fastest, after $(\tau-\tau_0)/\tau_0>2.2$, the chiral restored phase disappears. While the system with small relaxation time expands slower, it takes longer time for the chiral restored phase to disappear. It appears  from the evolution of mass (Fig.~\ref{gub_mass}) and the phase boundary (Fig.~\ref{gub_boundary}) that the various relaxation time does not have obvious influence on the quark mass. Since the order parameter describes the long-range correlation and the overall property of the system, while the collision is the local process in a system and is a short range correlation, the collision does not have big impact on the global symmetry of the system.

\subsection{Kink in velocity}
When solving together the transport equation and the gap equation, kinks in the velocity $u_\tau$ and $u_\rho$ are discovered. The velocity along $\rho$-direction $v_\rho \equiv u_\rho/u_\tau$ for different current mass and relaxation time are also presented in Fig.~\ref{gub_mass}. As we have mentioned above, the quark mass enters the transport equation through both the energy $E_p$ and the force term $\boldsymbol{\nabla} E_p\cdot\boldsymbol{\nabla_p} f$. The gradient of the field energy acts as a continuous force on the quarks, describing the interaction between the quarks and the mean fields. This interaction changes the quarks' momenta. Although this force term is also considered in the simulations such as test particle method~\cite{Abada:1994mf, Abada:1996bw,vanHees:2013qla,Meistrenko:2013yya,Wesp:2017tze}, its influence on the phase transition and the expansion has not been carefully investigated. For a chiral phase transition at low density, it is either a crossover or a second order phase transition, in both case the order parameter changes continuously. In comparison, the gradient of order parameter diverges at a second order phase transition, hence the force could be extremely strong.

\begin{figure}[H]\centering
\includegraphics[width=0.35\textwidth]{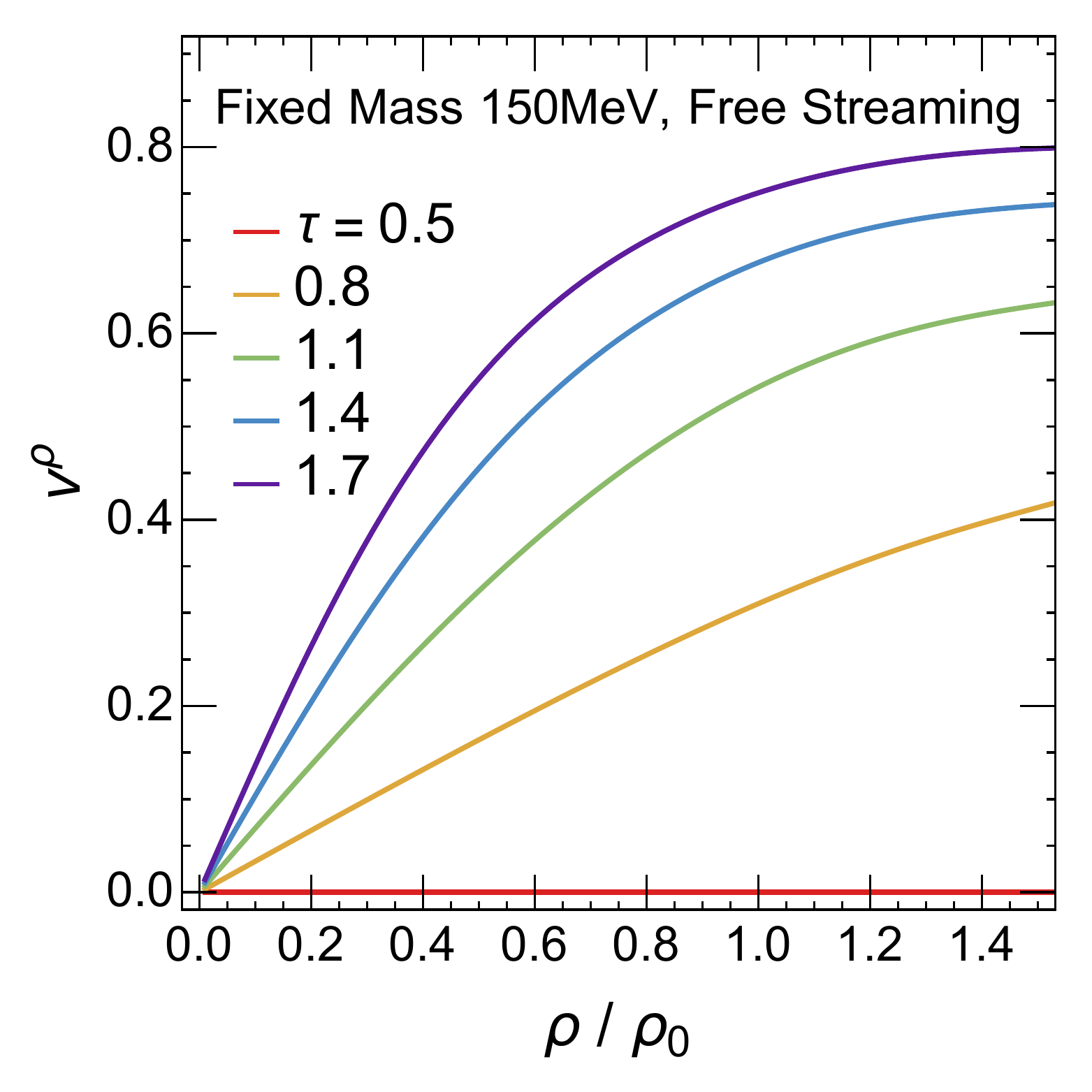}
\caption{Velocity in $\rho$-direction of the expansion with constant mass $m_\psi(\tau,\rho)=150$~MeV and infinite relaxation time. The lines are rainbow-colored represents different time in the evolution.}
\label{gub_mass_const}
\end{figure}

We first present the velocity $v_\rho$ in the scenario of constant mass so as to illustrate the influence of the phase transition on the evolution of the system, see Fig.~\ref{gub_mass_const}. We take constant homogeneous mass $m_\psi(\tau,\rho)=150$MeV and solve only the transport equation for a free streaming system, the force term vanishes since $\partial_\tau m=0$ and $\partial_\rho m=0$. In this scenario, the kink does not appear, indicating the kink arises from the inhomogeneous mass distribution and thus the force term.
\begin{figure}[H]\centering
\includegraphics[width=0.6\textwidth]{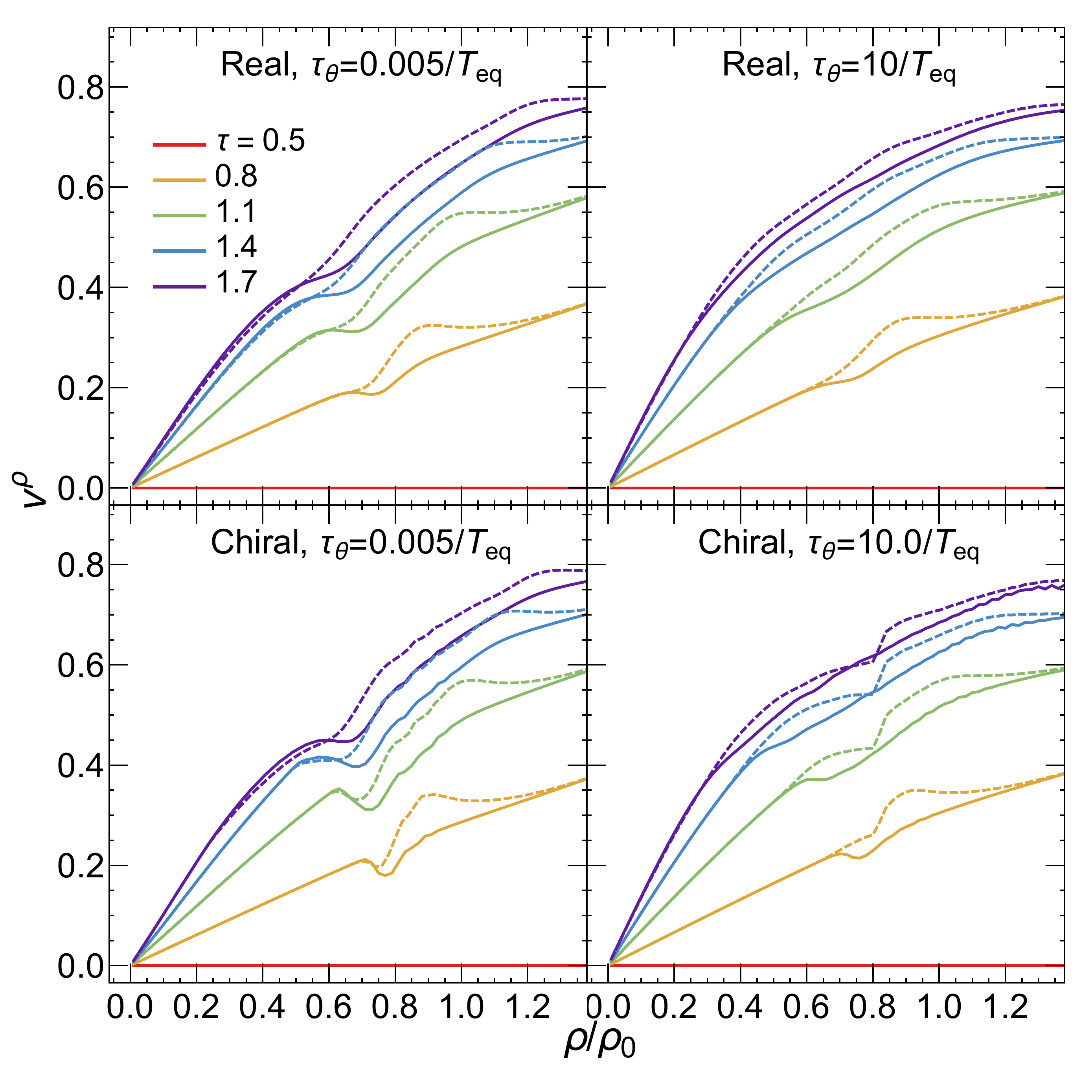}
\caption{The transverse velocity $v_\rho$ for real case and chiral limit, with different relaxation time. The solid lines are transverse velocity $v_\rho$ obtained by self-consistent solution. The dashed lines are transverse velocity $v_\rho$ obtained by solving together the transport equation and gap equation, but ignore the force terms in the transport equation.}
\label{gub_noforce}
\end{figure}
In order to further illustrate the influence of the force term, we now concurrently solve the transport equation and the gap equation, but removes the force terms in the transport equation by fixing $\partial_\tau m=0$ and $\partial_\rho m=0$. The evolution of mass and thermodynamic quantities has no obvious difference compared to those of self-consistent solutions, however, the velocity $v_\rho$ is quite different, which is presented as dashed lines in Fig.~\ref{gub_noforce}, with the solid lines are the velocity of self-consistent solution. As shown in each figures Fig.~\ref{gub_noforce}, when the force terms are ignored, the velocity has a bump around the phase transition. Namely if there is phase transition but no force term, the quark near the phase boundary are accelerated. In comparison, when the force term is considered, the quark near the phase boundary are slowed down. The appearance of kinks is closely related to the phase transition and the spatial distribution of the quark mass. The velocity kinks are more obvious in scenarios with small current mass or small relaxation time. This phenomenon can be understood as follows, since the force term is related to the susceptibility as well as the gradient of temperature $\boldsymbol{\nabla} m=(\mathrm{d}m/\mathrm{d}T)\boldsymbol{\nabla} T$, the phase transition of the chiral limit has divergent susceptibility while that of real case is finite, thus the kinks in the chiral limit panels are more obvious compared with those in real case panel. The expansion starts with equilibrium distribution and gradually becomes out-of-equilibrium due to the huge pressure gradient. With large relaxation time, the system spends longer time in the out-of-equilibrium state, the critical effect is further washed out. 

\subsection{Distribution function}
The phase transition hypersurface affects the motion of quarks around it, and has an influence on the $p_T$ spectrum of the quarks in the system. This can be directly revealed from the distribution function. Under the given symmetry in section III, the distribution function is defined on a $4+1$d phase space, $(\rho,p_\perp,\theta,\xi)$ as well as $\tau$, where $\theta$ is the angle between the transverse coordinate $\rho$ and the transverse momentum $p_T$. The $\theta$-dependence has been expanded to a series of Fourier coefficients. The zeroth component $a_0$, which is obtained by integrating the distribution function $f(\rho, p_T, \theta, \xi)$ over $\theta$ is related to the density distribution. While the first component $a_1$ obtained by integrating $f(\rho, p_T, \theta, \xi)\cos\theta$ over $\theta$ gives hint to the direction of the particle velocity. 

\begin{figure}[H]\centering
\includegraphics[width=0.35\textwidth]{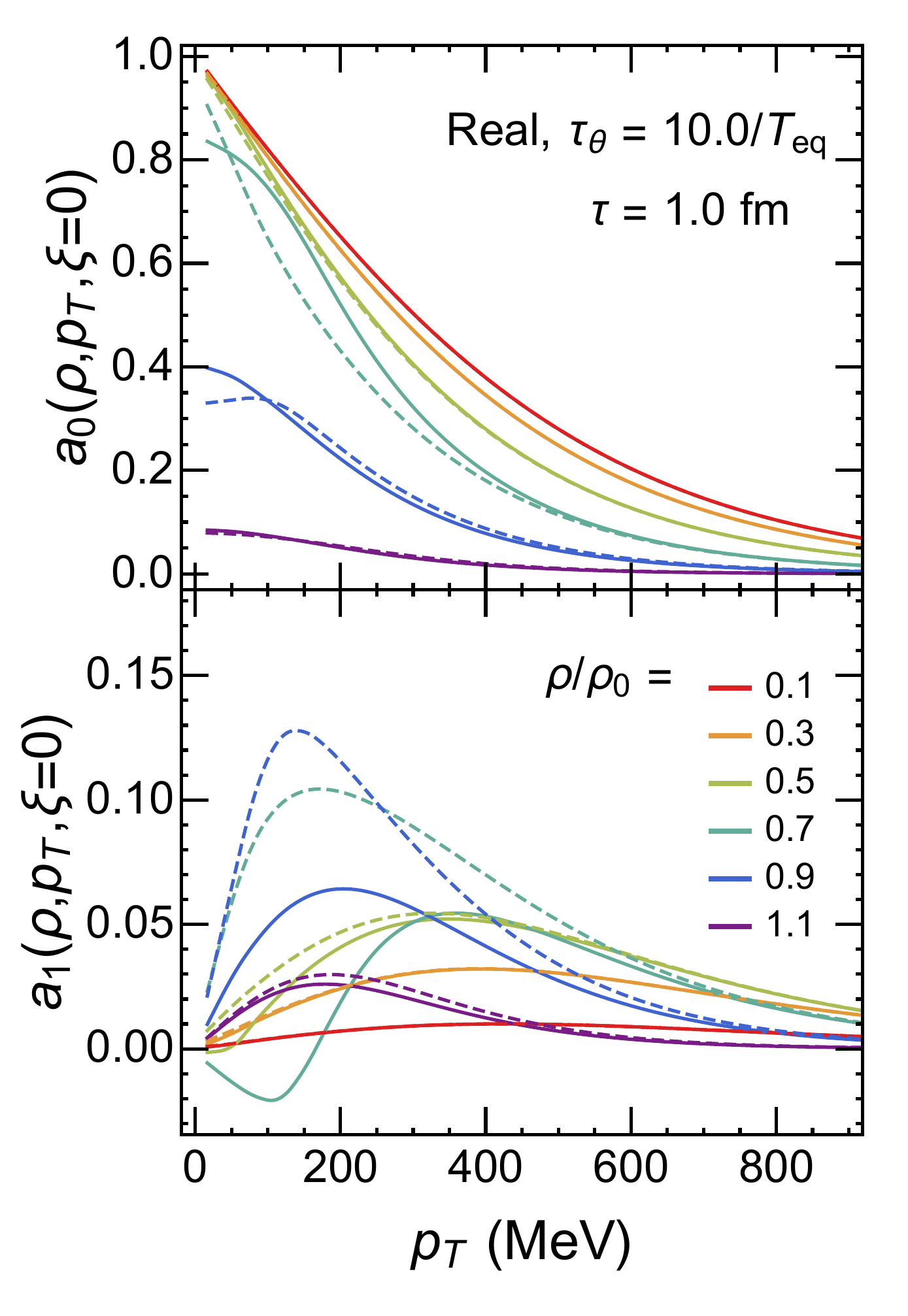}\qquad
\includegraphics[width=0.35\textwidth]{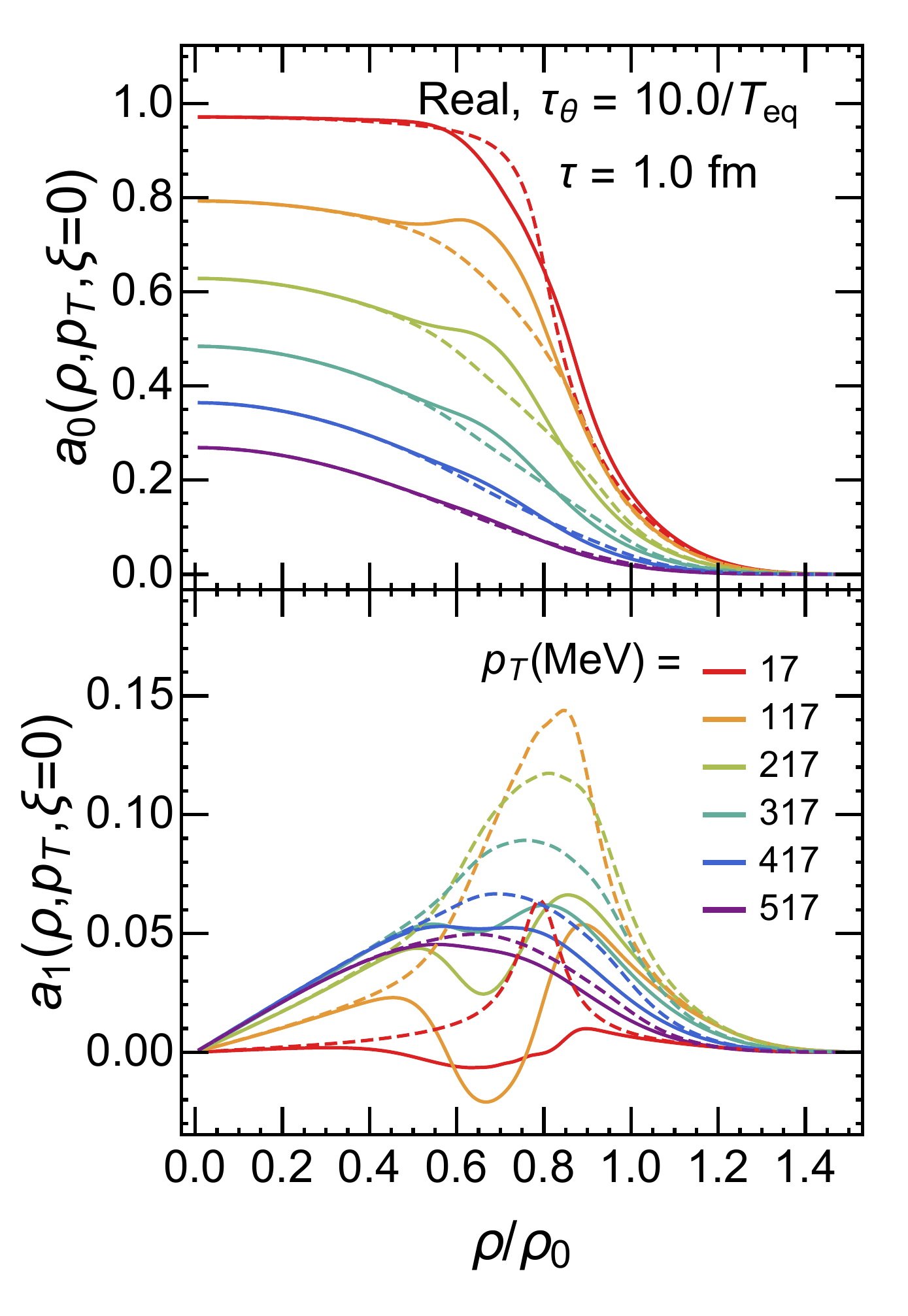}
\caption{Zeroth and the first Fourier components in the case the collisions are weak $\tau_\theta=10/T_\mathrm{eq}$, the solid lines are the distribution function from the self-consistent solution, the dashed lines are the distribution function when the force term is ignored. The left panel are $a_0$ and $a_1$ as functions of transverse momentum $p_T$ at various given transverse coordinates $\rho/\rho_0$, the right panel shows $a_0$ and $a_1$ as functions of transverse coordinates $\rho/\rho_0$ at various given transverse momentum $p_T$.}
\label{distribution1}
\end{figure}

In Fig.~\ref{distribution1}, we present the zeroth and the first Fourier components in the scenario with weak collisions, namely with large relaxation time $\tau_\theta=10/T_\mathrm{eq}$. 
The solid lines correspond the self-consistent distribution function, while the dashed lines are the distribution function where the force term is ignored. 
Coefficients $a_0$ and $a_1$ as a function of transverse momentum $p_T$ at various given transverse coordinates $\rho/\rho_0$ are presented in the left panel. 
The right panel shows the coefficients $a_0$ and $a_1$ as a function of transverse coordinates $\rho/\rho_0$ at various given transverse momentum $p_T$. 

At some evolution time $\tau=1.0$ fm or $(\tau-\tau_0)/\tau_0=1$, the phase transition takes place around the position $\rho/\rho_0\sim 0.7$, which is also the location of the kink in the velocity, see Fig.~\ref{gub_mass}. 
From the left panel of Fig.~\ref{distribution1}, in the self-consistent solution (solid lines), the distribution function $a_1$ is negative at low $p_T$ for $\rho/\rho_0$ around 0.6 to 0.8, which means the particles with low momentum are bounced back by the phase transition ``wall''; while the dashed lines reveals that without force term, the particles cannot see the ``wall''. 
This effect is also obvious from the $a_1$ in the right panel, for small momentums, the distribution function changes sign around $\rho/\rho_0=0.7$; for large momentums, the distribution function stays positive but has smaller values. This indicates that the particles with small momentum are bounced back by the phase transition wall around $\rho/\rho_0=0.7$, particles with large momentum go through the ``wall'', but have been slowed down. 
The integral of $a_0(\rho, p_T, \xi)$ over $p_T$ at fixed $\rho$ corresponds to the number density of particles somewhere in the transverse plane, while integral of $a_0(\rho, p_T, \xi)$ over $\rho$ at fixed $p_T$ is the number density of particles of some fixed momentum. It can be observed from the left panel, whether the force term is ignored or not, the number density away from the ``wall'' ($\rho/\rho_0=0.7$) is similar, while the force term would collect more particles around the ``wall''. From the right panel, there are more low momentum particles kept inside the wall because of the force term.
\begin{figure}[H]\centering
\includegraphics[width=0.35\textwidth]{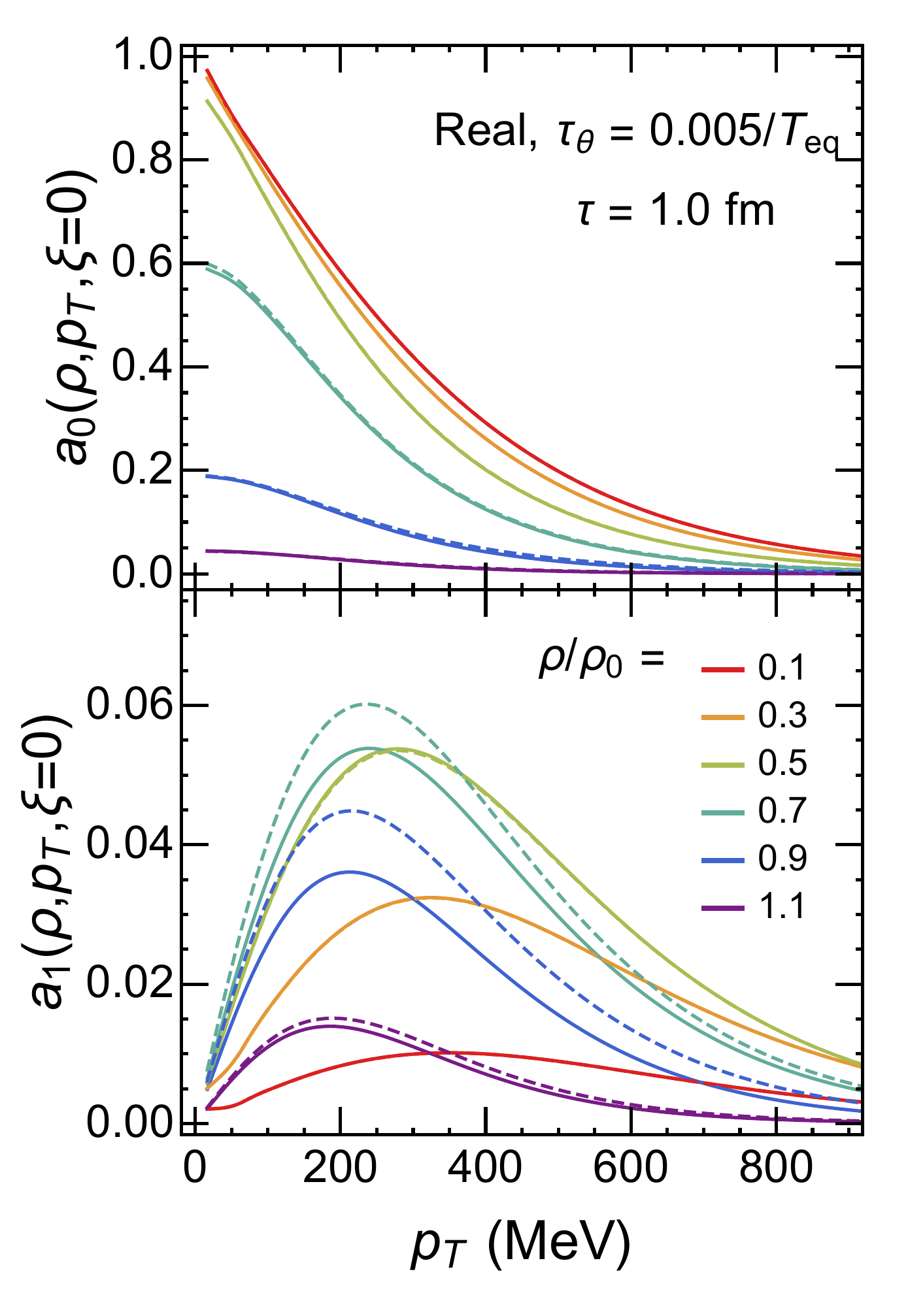}\qquad
\includegraphics[width=0.35\textwidth]{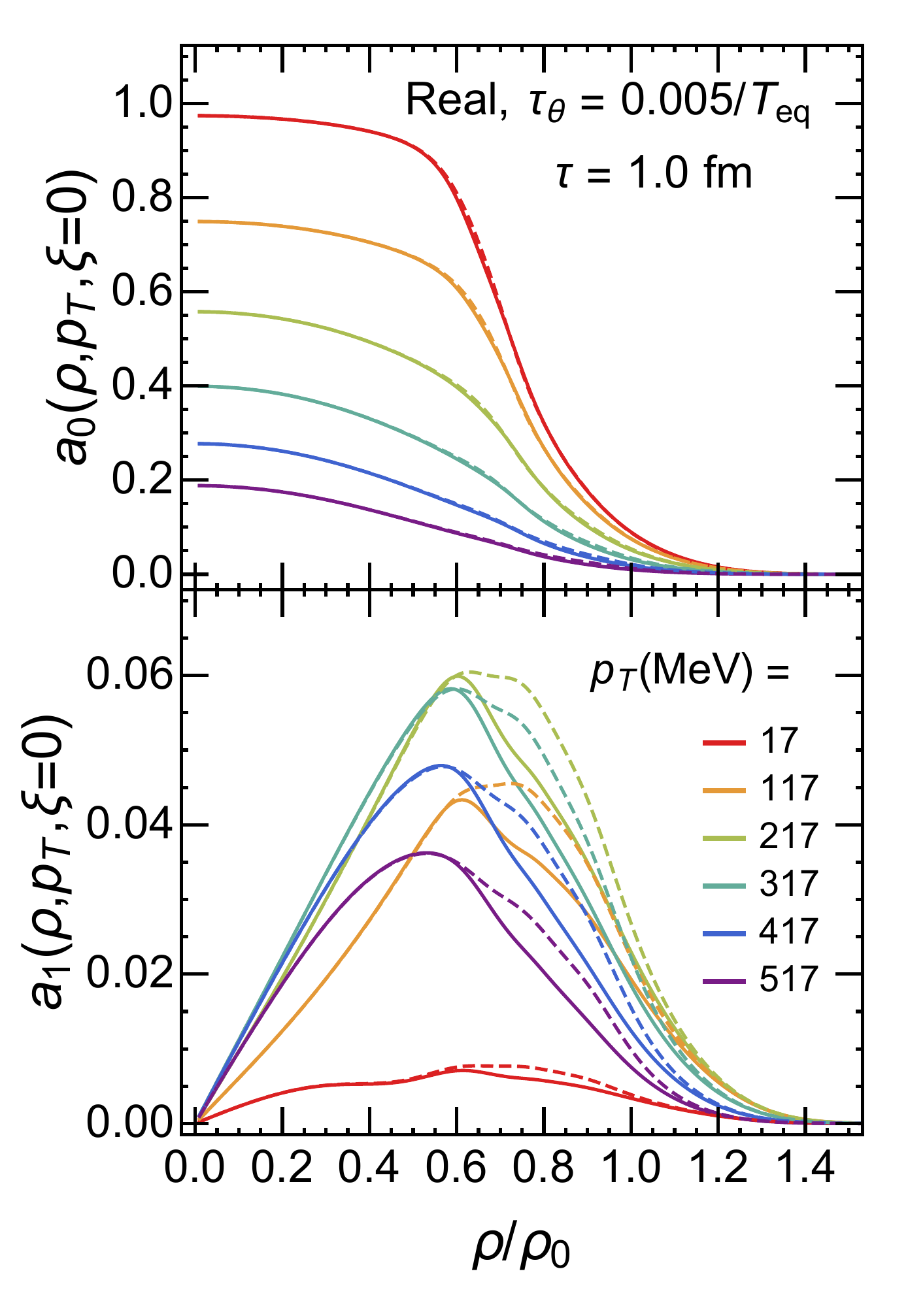}
\caption{Zeroth and the first Fourier components in the case the collisions are strong $\tau_\theta=0.005/T_\mathrm{eq}$, the solid lines are the distribution function from the self-consistent solution, the dashed lines are the distribution function when the force term is ignored. The left panel shows $a_0$ and $a_1$ as functions of transverse momentum $p_T$ at various given transverse coordinates $\rho/\rho_0$, the right panel shows $a_0$ and $a_1$ as functions of transverse coordinates $\rho/\rho_0$ at various given transverse momentum $p_T$.}
\label{distribution2}
\end{figure}

As is presented in Fig.~\ref{distribution2}, the effect of the phase transition ``wall'' is smoothed by the collision. Since the particles are relaxed into thermal distribution, the direction of a single particle is aligned along the collective velocity. With strong collisions, the influence of the force term on number density almost disappear, giving the same $a_0$ whether the force term is taken into consideration. The first order components $a_1$ is still affected by the force term. Inside the ``wall'' ($\rho/\rho_0<0.7$), the force term does not has any influence. Outside the ``wall'' ($\rho/\rho_0>0.7$), the distribution $a_1$ is smaller when the force term is considered, which means the particles are slowed down by the force term. 

\section{Summary}
Quite different from the equilibrium thermodynamics, the realistic chiral phase transition in heavy ion collision takes place in a highly inhomogeneous, fast evolving dynamical system. This requires the understanding of the chiral phase transition in an expanding, out-of-equilibrium system. 

In this work, we investigate the evolution of an expanding quark-antiquark system with self-consistently taking into account the dynamical quark constituent mass. In order to reduce the dimension of the phase space, we consider a longitudinal boost invariant and transversal rotational symmetric system, which is a good approximation for ultra-central heavy ion collisions. In the numerical process, both Vlasov and gap equations are solved concurrently, giving a self-consistent evolution of both the quark-antiquark distribution function and the quark constituent mass. The spacetime-dependent constituent mass serves as the chiral order parameter and affects the evolution of the quark distribution function through the force term. In order to investigate the off-equilibrium effects, we introduce relaxation time approximation for the collision term, and compare the local equilibrium and out-of-equilibrium result by considering small and large relaxation time. 

The evolution of the quark mass illustrates the chiral phase transition, and defines the phase diagram in the space-time. A kink in the transverse velocity in observed around the phase transition boundary, which appears because the large force term around the phase transition. The kink is more obvious for smaller current mass and smaller relaxation time, which means that the crossover and non-equilibrium effect tends to smooth out the kink. The influence of the phase transition hypersurface is further investigated by directly analyzing the distribution function. It is observed that, the phase transition wall would bounce back the low momentum particles, thus gives a kink in the velocity and may enhance the low $p_T$ part of the momentum spectrum. The kink indicates that the parton decelerates when propagating from hot to cold region because of the increase in its effective mass. This effect may have impact on the produced hadrons and other observables in experiment, such as reducing the mean $p_T$ and the enhancement of particle production yield, which will be investigated in future work.

\noindent  {\bf Acknowledgement}: The work is supported by the NSFC Grant Number 11890712. ZyW is supported by the Postdoctoral Innovative Talent Support Program of China, and SS by the Natural Sciences and Engineering Research Council of Canada.
\bibliographystyle{elsarticle-num}
\bibliography{ref}

\end{document}